\documentclass[%
 reprint,
superscriptaddress,
preprintnumbers,
 amsmath,
 amssymb,
 aps,
 prl,
]{revtex4-2}

\usepackage{graphicx}
\usepackage{dcolumn}
\usepackage{bm}
\usepackage[mathlines]{lineno}
\usepackage[usenames,dvipsnames]{color} 
\usepackage{soul}

\newcommand{\Cs}{CsV\textsubscript{3}Sb\textsubscript{5}}
\newcommand{\Tc}{\textit{T}\textsubscript{c}}
\newcommand{\Tcdw}{\textit{T}\textsubscript{CDW}}

\begin{document}


\title{Colossal \textit{c}-axis response and lack of rotational symmetry breaking \\within the kagome plane of the \Cs~superconductor}

\author{Mehdi Frachet}
 \email{mehdi.frachet@gmail.com}
 \affiliation{Institute for Quantum Materials and Technologies, Karlsruhe Institute of Technology, D-76021 Karlsruhe, Germany}

\author{Liran Wang}
 \affiliation{Institute for Quantum Materials and Technologies, Karlsruhe Institute of Technology, D-76021 Karlsruhe, Germany}

\author{Wei Xia}
 \affiliation{School of Physical Science and Technology, ShanghaiTech University, Shanghai 201210, China}
 \affiliation{ShanghaiTech Laboratory for Topological Physics, Shanghai 201210, China}

\author{Yanfeng Guo}
 \affiliation{School of Physical Science and Technology, ShanghaiTech University, Shanghai 201210, China}
 \affiliation{ShanghaiTech Laboratory for Topological Physics, Shanghai 201210, China}
 
 \author{Mingquan He}
 \affiliation{Low Temperature Physics Laboratory, College of Physics \& Center of Quantum Materials and Devices,
Chongqing University, Chongqing 401331, China}

\author{Nour Maraytta}
 \affiliation{Institute for Quantum Materials and Technologies, Karlsruhe Institute of Technology, D-76021 Karlsruhe, Germany}

\author{Rolf Heid}
 \affiliation{Institute for Quantum Materials and Technologies, Karlsruhe Institute of Technology, D-76021 Karlsruhe, Germany}

\author{Amir-Abbas Haghighirad}
 \affiliation{Institute for Quantum Materials and Technologies, Karlsruhe Institute of Technology, D-76021 Karlsruhe, Germany}

\author{Michael Merz}
 \affiliation{Institute for Quantum Materials and Technologies, Karlsruhe Institute of Technology, D-76021 Karlsruhe, Germany}
 \affiliation{Karlsruhe Nano Micro Facility (KNMFi), Karlsruhe Institute of Technology, 76344 Eggenstein-Leopoldshafen\\}

\author{Christoph Meingast}
 \email{christoph.meingast@kit.edu}
 \affiliation{Institute for Quantum Materials and Technologies, Karlsruhe Institute of Technology, D-76021 Karlsruhe, Germany}

\author{Frédéric Hardy}
 \email{frederic.hardy@kit.edu}
 \affiliation{Institute for Quantum Materials and Technologies, Karlsruhe Institute of Technology, D-76021 Karlsruhe, Germany}


\date{\today}

\begin{abstract}
The kagome materials AV$_3$Sb$_5$ (A = K, Rb, Cs) host an intriguing interplay between unconventional superconductivity and charge-density-waves. Here, we investigate CsV$_3$Sb$_5$ by combining high-resolution thermal-expansion, heat-capacity and electrical resistance under strain measurements. We directly unveil that the superconducting and charge-ordered states strongly compete, and that this competition is dramatically influenced by tuning the crystallographic $c$-axis. In addition, we report the absence of additional bulk phase transitions within the charge-ordered state, notably associated with rotational symmetry-breaking within the kagome planes. This suggests that any breaking of the $C_{\rm{6}}$ invariance occurs via different stacking of $C_6$-symmetric kagome patterns. Finally, we find that the charge-density-wave phase exhibits an enhanced $A_{\rm{1g}}$-symmetric elastoresistance coefficient, whose large increase at low temperature is driven by electronic degrees of freedom.

\end{abstract}

\maketitle



The unique electronic band structure of delocalized electrons in kagome lattices features Dirac points, flat bands, and multiple van Hove singularities (vHS) close to the Fermi level \citep{Hu_NatComm_2022}. Theoretical studies of kagome lattices demonstrate that the large density of state near van Hove filling can promote various exotic electronic orders, including charge-bond order, chiral charge-density-wave, orbital-current order, and superconducting states of various gap symmetries \citep{Kiesel_PRL_2013, Denner_PRL_2021,Shen_PRB_2013}. 

In this context, the family of recently discovered kagome metals AV$_3$Sb$_5$ (A = K, Rb, Cs), crystallizing in the $P$6/$mmm$ hexagonal space-group with perfect vanadium kagome networks, has emerged as an exciting realization of such physics with nontrivial topological properties, unconventional superconductivity and  intertwined symmetry-broken states \citep{Ortiz_PRM_2019, Ortiz_PRL_2020}. Experimentally, two electronic instabilities are well established in all AV$_3$Sb$_5$, {\it i.e.} a charge-density wave (CDW) below $T_{\rm{CDW}}$ $\approx$ 100 K, and bulk superconductivity (SC) that reaches $\Tc \approx 2.5$ K in CsV$_3$Sb$_5$. The CDW state features a triple-\textbf{q}-modulation with wave-vector connecting the three inequivalent sublattices and the corresponding M saddle-points vHS. Below \Tcdw, the translational symmetry of the crystal lattice is broken, but the $c-$axis periodicity remains highly debated as, {\it e.g.}, $2\times2
\times2$ \citep{Liang_PRX_2021, Tan_PRL_2021} and $2\times2\times4$ \citep{Ortiz_PRX_2021} superstructures, or a combination thereof \citep{Xiao_PRX_2023, Stahl_PRB_2022}, are reported.


The fate of the six-fold rotational invariance of the hexagonal lattice is controversial. Several experiments including x-ray diffraction (XRD), nuclear magnetic resonance (NMR) and scanning tunneling microscopy (STM) point to a lowering to $C_{2}$ rotational symmetry \citep{Xu_NatPhys_2022, Xiang_NatComm_2021, Nie_Nature_2022, Stahl_PRB_2022}. In addition, in \Cs, measurements of the electrical resistance under strain, namely elastoresistance, have been interpreted as an evidence for a growing electronic nematic susceptibility within the $E_{2g}$ ($x^2-y^2$) symmetry channel, ultimately leading to an ordered nematic state at $T_{\rm{nem}} =35~$K \citep{Nie_Nature_2022, Sur_npjQM_2023}. However, different experiments suggest different critical temperatures for the $C_{6}$-symmetry breaking, ranging from $T_{\rm{nem}}$ to \Tcdw, although no thermodynamic evidence for such transition has been found. Further, conflicting results regarding a possible time-reversal symmetry-breaking at \Tcdw~ were reported \citep{Saykin_PRL_2023, Hu_Arxiv_2023, Xu_Nature_2022, Mielke_Nature_2022}, such that it remains unsettled whether AV$_3$Sb$_5$ could be the hosts of, {\it e.g.}, a long sought loop current order \citep{Xu_Nature_2022, Mielke_Nature_2022, Mortensen_PRB_2022}.

Although a conventional mechanism is unable to explain the superconducting state of AV$_3$Sb$_5$ \citep{Tan_PRL_2021}, its nature remains unsettled. No consensus has been reached concerning the gap symmetry and the existence of gap nodes \citep{Zhao_Arxiv_2021,Duan_ScienceChina_2021, Gupta_npjQM_2022}. Further, it has been proposed that the SC and CDW states conspire to form a pair-density-wave \citep{Chen_Nature_2021}, and, importantly, that electronic nematicity plays a key role in the mechanism of superconductivity in {\it e.g.} Cs(V$_{1-x}$Ti$_x$)$_3$Sb$_5$ \citep{Sur_npjQM_2023}.

In this Letter, we use a powerful combination of bulk thermodynamic measurements, including high-resolution thermal-expansion and heat-capacity, with elastoresistance on \Cs~single crystals from two different sources to gain further insights into the CDW state and its connection with superconductivity.
Our results directly demonstrate (i) a strong competition between the CDW and the SC states, dramatically influenced by $c-$axis tuning, (ii) the absence of an orthorhombic distortion for $T\leq$ \Tcdw, implying that either the CDW does not break 6-fold symmetry or that rotational symmetry-breaking is decoupled from anisotropic strain and (iii) that the CDW is characterized by a strongly enhanced $A_{\rm{1g}}$-symmetric elastoresistance, which further increases with decreasing temperature.

Single crystals of \Cs~ were grown in Shanghai (batch A) and Karlsruhe (batch B) by the flux method and characterized by x-ray diffraction and energy-dispersive x-ray analysis (see Supplemental Material). In-plane thermal-expansion measurements were carried out using a home-built high-resolution capacitive dilatometer on a large single crystal from batch A. Because of the large aspect ratio, $c$-axis measurements were performed using a stack of 20 smaller crystals from batch B, glued together with GE7031 varnish to a thickness of $\approx$ 2 mm. Elastoresistivity measurements were carried out by gluing samples from batch B on a piezoelectric stack (Pst 150/5x5/7 from Piezomechanik) using DevCon 5mn
2-components epoxy (Part. No. X0039) as described in Ref. \citep{Frachet_npjQM_2022}. To extract the symmetry-resolved elastoresistance coefficients we assumed a temperature-independent Poisson ratio $\nu_p= -\epsilon_{yy}/\epsilon_{xx} \approx 0.43$ \citep{Kuo_PRB_2013} for the piezoelectric stack ($x$ is the poling direction). Heat-capacity measurements $C$(T) were made on the same sample from batch A in a Physical Property Measurement System from Quantum Design.


\begin{figure}
\centering
\includegraphics[width=8.6cm]{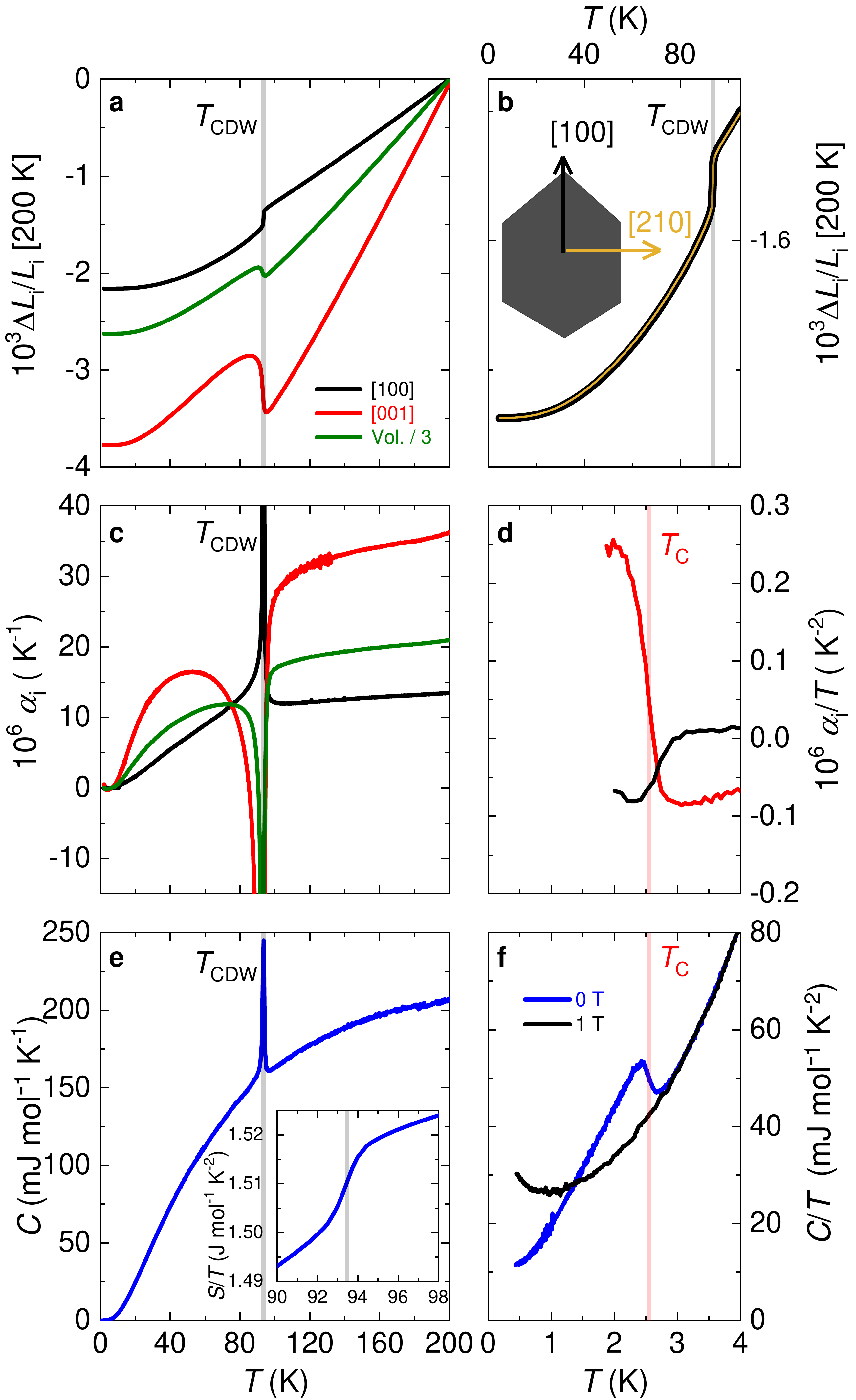}
\caption{\textbf{(a)} Relative length changes, $\Delta L/L$, along the hexagonal directions and corresponding volume change as a function of temperature. The black vertical dashed line indicates \Tcdw. \textbf{(b)} Comparison of $\Delta L_{i}/L_{i}$ measured along the orthogonal [100] and [210] hexagonal directions. \textbf{(c)} Corresponding thermal-expansion coefficient, $\alpha =  1/L \left(dL/dT\right)$. \textbf{(d)} shows the thermal-expansion data for $T<$ 4K on a magnified view. \textbf{(e)} Heat capacity $C(T)$ showing the first-order transition at \Tcdw~and the corresponding entropy discontinuity (inset). \textbf{(f)} shows the superconducting transition in the specific heat on an extended view. The red vertical dashed line indicates \Tc.  }
\label{Fig1}
\end{figure}


Figure \ref{Fig1}(a) shows the relative length changes, $\Delta L/ L$, of our sample as a function of temperature. A clear first-order discontinuity, for both [100] and [001] crystal axes (in the following we use $P6/mmm$ space group notations), accompanied by a large peak in the specific heat (see Fig.\ref{Fig1}(e)) are observed at \Tcdw $\approx$ 93.5 K, marking the transition to the CDW state. Although the CDW transition is clearly of first-order, we also observe significant fluctuation effects both above and below the transition in an appropriate Grüneisen parameter (see Supplemental Materials). At lower temperature, second-order discontinuities are clearly resolved at \Tc = 2.5 K in both thermal-expansion coefficient, $\alpha_{i}(T) = 1/L_{i} \left(dL_{i}/dT\right)$ with $i = \{[100], [001]\}$, and  $C$(T), as illustrated in Figs \ref{Fig1}(d) and \ref{Fig1}(f), respectively. However, we find no evidence of a phase transition around 60 K (see Figs \ref{Fig1}(a) and \ref{Fig1}(c)), especially in the $c$-axis thermal expansion, where sharp changes in the intensity of the superstructure reflections, accompanying the change of interlayer ordering, were observed by XRD \citep{Stahl_PRB_2022}. This is rather surprising since band folding, resulting from a changing superstructure, is expected to substantially modify the Fermi surface and therefore the electronic entropy. Yet, no signature of this transition is resolved in either $C(T)$ or $\alpha(T)$, which measure the T- and p-derivative of the entropy, respectively.

Our measurements, however, clearly demonstrate a huge dependence of both CDW and SC on uniaxial pressure. Table \ref{table_thermo} summarizes the initial uniaxial- and hydrostatic-pressure dependences of \Tcdw~and \Tc~inferred from the application of the Clausius-Clapeyron and Ehrenfest relations, respectively. The largest effect is found for $c$-axis where $d\Tcdw/dp_{c}$ amounts to $\approx$ -120 K GPa\textsuperscript{-1}. This demonstrates that the large hydrostatic-pressure sensitivity of the CDW instability of \Cs~\citep{Zheng_Nature_2022, Chen_PRL_2021} originates predominantly from $c-$axis stress and highlights the importance of the apical Sb-bonds and Sb-derived bands \citep{Turan_PRB_2023, Ritz_Arxiv_2023}. This is equally true for SC which also exhibits large uniaxial pressure dependences but opposite in sign, confirming that both orders are competing for the same electronic states \citep{Nie_PRB_2021}.

Remarkably, the relative change of \Tc~with $c-$axis stress, $1/\Tc~\left(d\Tc /dp_{\rm{c}}\right)$, is roughly a factor 4 greater than that of \Tcdw. Furthermore, the relative changes of the Sommerfeld coefficient $\gamma$ and \Tc~with $c-$axis pressure are both positive. Thus, the increase of \Tc~under $c-$axis pressure is likely explained by an increase in the density of states due to the reduction of the CDW gap. Interestingly, this correlates with the convergence of M saddle-points vHS toward the Fermi level \citep{Consiglio_PRB_2022}.


Interestingly, we find no evidence for additional phase transition, in contrast to the reports of $C_6$-symmetry breaking at $T_{\rm{nem}} = 35$ K by elastoresistance and NMR \citep{Sur_npjQM_2023, Nie_Nature_2022}, or directly below $T_{\rm{CDW}}$ by x-ray diffraction \citep{Stahl_PRB_2022}. Such symmetry-breaking should be detected using our high-resolution capacitive dilatometer by comparing the strains measured along the [100] and the orthogonal [210] directions, as has been demonstrated for several Fe-based superconductors \citep{Bohmer_NatComm_2015, Wang_PRB_93}. This is because our spring-loaded dilatometer exerts a non-negligible stress along the measurement direction. Thus, for a measurement along the hexagonal [210] direction, the population of possible structural domains with the shorter orthorhombic axis should be favored, resulting in an in-situ detwinning of the sample below \Tcdw, if the crystal symmetry were lowered. On the other hand, the twin population would remain unaffected by the applied force for measurements along the [100] direction, which probe a mixture of both orthorhombic axes. As illustrated in Figure \ref{Fig1}(b) we find no discernible difference between the two measurements suggesting either that the CDW does not break $C_6$ symmetry or that its broken rotational symmetry is decoupled from anisotropic strain.

\begin{table}[]
    \centering
    \begin{tabular}{c|c|c||c}
     & a & c & Volume  \\
    \hline
    \hline
    $d\ln{\Tc}/{dp_i}$ [GPa\textsuperscript{-1}]  & -1.3  &  +4.7  & +2.1 \\
    \hline
    {${d\ln{\Tcdw}}/{dp_i}$} [GPa\textsuperscript{-1}]  & +0.24 & -1.3 & -0.81   \\
    \hline
    \textbf{${d\ln{\gamma}}/{dp_i}$} [GPa\textsuperscript{-1}]  & -0.06  &  +0.73  & +0.58\\
        

    \end{tabular}
    
    \caption{Relative variations of $\Tc$, $\Tcdw$, and the Sommerfeld coefficient $\gamma$ with uniaxial pressure, calculated using the Ehrenfest, Clausius-Clapeyron and Maxwell relations, respectively.} 
    
    \label{table_thermo}
    
\end{table}

\begin{figure}
\centering
\includegraphics[width=8.6cm]{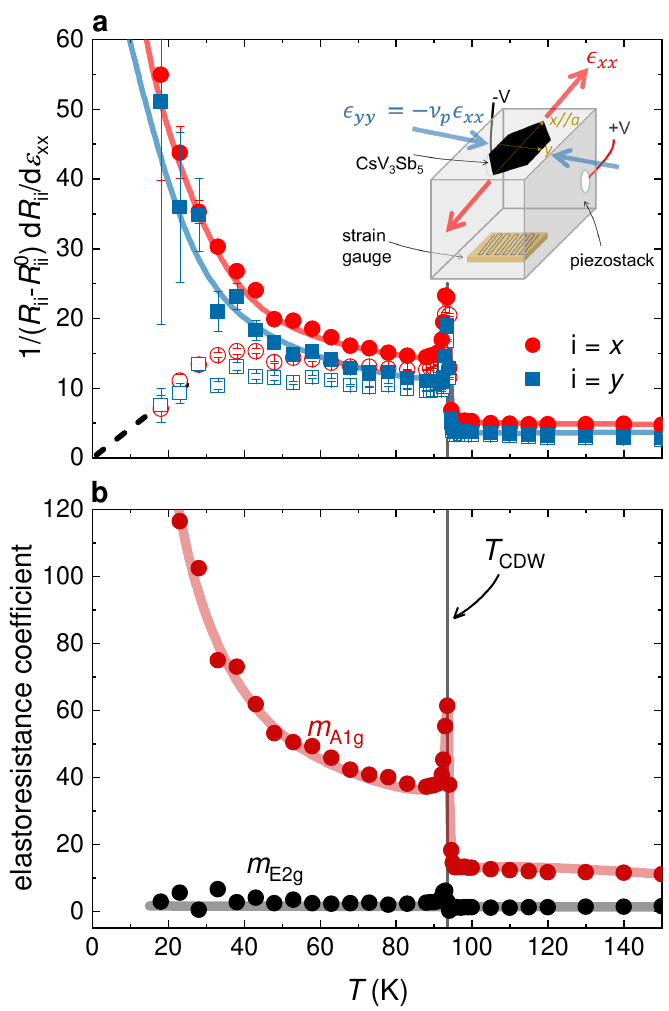}
\caption{\textbf{(a)} Linear slopes of the resistance versus strain curves, $1/\left(R_{ii}-R_0\right) \left(dR_{ii}/d\epsilon_{xx}\right)$, in the longitudinal ($i=x$, red symbols) and transverse ($i=y$, blue symbols) channels. The empty symbols correspond to resistance variation relative to the total resistance (including the residual term), $1/R_{\rm{ii}}\left(dR_{\rm{ii}}/d\epsilon_{\rm{xx}}\right)$, as discussed in earlier works \citep{Nie_Nature_2022, Sur_npjQM_2023}. The inset shows a sketch of the experimental setup with the sample (black) glued on the top side of a piezoelectric stack (gray) with the crystallographic $a$-axis aligned along the poling direction ($x$). \textbf{(b)} Corresponding $E_{\rm{2g}}$ and $A_{\rm{1g}}$ symmetry-resolved elastoresistance coefficients. The vertical line indicates \Tcdw.}
\label{Fig2}
\end{figure}


The lack of evidence for $C_6$-symmetry breaking in our thermal expansion measurements motivates a closer inspection of the changes of electrical resistance, $\Delta R_{ii} = R_{ii}(\epsilon_{\rm{xx}})-R_{ii}(\epsilon_{\rm{xx}}=0)$ with $i = \{\rm{x}, \rm{y}\}$, in response to applied strain $\epsilon_{\rm{xx}}$. Here, we induce a small symmetry-breaking strain to our crystals using the technique introduced in Ref.\citep{Chu_Science_2012}, as depicted in the inset of Fig. \ref{Fig2}a. With the [100] axis of \Cs~ aligned with the piezoelectric poling direction $\rm{x}$, we extract,

\begin{equation}
    \left(\dfrac{\Delta R}{R-R^{0}}\right)_{\rm{xx}}-\left(\dfrac{\Delta R}{R-R^{0}}\right)_{\rm{yy}} = m_{\rm{E_{2g}}}  (\epsilon_{\rm{xx}}-\epsilon_{\rm{yy}}),
    \label{eq1}
\end{equation}
 
\begin{equation}
    \left(\dfrac{\Delta R}{R-R^{0}}\right)_{\rm{xx}}+\left(\dfrac{\Delta R}{R-R^{0}}\right)_{\rm{yy}} = m_{\rm{A_{1g}}}  (\epsilon_{\rm{xx}}+\epsilon_{\rm{yy}}),
    \label{eq2}
\end{equation}
where $R_{ii}^0$ is the $T\rightarrow$ 0 residual resistance and $R_{\rm{xx}}$ and $R_{\rm{yy}}$ correspond to resistance measurements along and transverse to the poling direction, {\it i.e.} parallel and perpendicular to the [100] hexagonal axis, respectively. Hereafter, we denote them longitudinal and transverse measurements, respectively. $m_{\rm{A_{1g}}}$ and $m_{\rm{E_{2g}}}$ represent the elastoresistance coefficients that transform according to the $A_{\rm{1g}}$ and $E_{\rm{2g}}$ irreducible representations of the $D_{\rm{6h}}$ point group, respectively (see Supplemental Material for details). Importantly, we normalize $\Delta R_{ii}$ by $\left(R_{ii}-R_{ii}^0\right)$ instead of $R_{ii}$ in order to obtain physically meaningful results at low temperatures, as discussed in Ref.\citep{Wiecki_NatComm_2021}.



In Fig. \ref{Fig2}, we report the results of our elastoresistance measurements on \Cs~ (see Supplemental Material for the raw data). The linear slopes of the resistance versus strain curves, $1/\left(R_{ii}-R_0\right) \left(dR_{ii}/d\epsilon_{xx}\right)$, are shown in Fig. \ref{Fig2}(a). For $T>T_{\rm{CDW}}$, the response to strain is weakly temperature dependent and amounts to $\approx 2-4$, as expected in any metals. At $T \approx T_{\rm{CDW}}$, a sharp peak is resolved in both directions. This peak, that has not been resolved in any AV$_{3}$Sb$_{5}$ \citep{Nie_Nature_2022,Sur_npjQM_2023}, is, however, naturally expected given the strong uniaxial pressure dependence of \Tcdw~ according to
    
\begin{equation}
    \left(\frac{d R}{d \epsilon_{\rm{xx}}}\right)_{T_{\rm{CDW}}} \approx   \left( \frac{\partial R }{\partial T} \right)_{T_{\rm{CDW}}}   \left( \frac{d T_{\rm{CDW}} }{d p_{a}}\right).
    \label{Eq_elastoPeak}
\end{equation}

The validity of Eq.(\ref{Eq_elastoPeak}) is provided by the essentially similar strain conditions achieved under uniaxial pressure and the in-plane biaxial and anisotropic strain induced by the piezostack (see details in Supplemental Material). Thus, a positive elastoresistance peak implies that $d T_{\rm{CDW}}/d p_{a} > 0 $, in agreement with our thermal-expansion measurements (see table \ref{table_thermo}) and previous direct uniaxial-stress experiments \citep{Nie_PRB_2021}. 

For $T \lesssim \Tcdw$, the elastoresistance does not turn back to a typical metallic value, but it is significantly enhanced \citep{Sur_npjQM_2023, Nie_Nature_2022} for both longitudinal and transverse channels, as illustrated in Fig.2(a). Hence, the electronic properties of \Cs~in the CDW state are mainly sensitive to a symmetry-preserving stress, in excellent accord with our thermodynamic results. In contrast to previous reports \citep{Nie_Nature_2022, Sur_npjQM_2023}, both our longitudinal and transverse measurements were carried out on the same sample, {\it i.e.} under similar strain-transmission conditions, which is crucial for extracting the symmetry-resolved elastoresistance coefficients shown in Fig.\ref{Fig2}(b). Specifically, the enhanced elastoresistance response within the CDW phase is totally dominated by the symmetry-preserving $A_{\rm{1g}}$ channel. The $E_{\rm{2g}}$ response, in contrast, is weak over the entire temperature range studied, in agreement with both the absence of in-plane $C_{6}$-symmetry breaking and the direct uniaxial strain measurements of Qian et al. \citep{Nie_PRB_2021}. In light of our thermodynamic results, the large $m_{\rm{A_{1g}}}$ likely originates from a dominant $c-$axis contribution.

Strikingly, the $A_{\rm{1g}}$-symmetric elastoresistance increases further in the CDW phase and reaches extremely high values at low temperature, with $m_{\rm{A_{1g}}}(20\rm{K})\approx 120$. This arises from the predominance of the electron-electron scattering at low temperature, because the electron-phonon scattering term to $\left(R_{xx}-R_{xx}^{0}\right)$ decreases faster than that of the electron-electron contribution, such that $m_{\rm{A_{1g}}}$ effectively increases. This is again in line with our thermodynamic data since this increased $m_{\rm{A_{1g}}}$ correlates with the decrease of $\gamma$ with [100] uniaxial pressure (see Table I) by virtue of the Kadowaki-Woods relation, $A\propto \gamma^2$, which relates $\gamma$ to the quadratic term of the resistivity in a Fermi liquid \citep{Wiecki_NatComm_2021}.        
Finally, no downturn of the elastoresistance is found below $T_{\rm{nem}}\approx$ 35 K \footnote{Note that our own measurements do also show a downturn below $\approx$ 35K when uncorrected for the residual resistivity. See empty symbols in Fig.2(a)}, strongly suggesting that its observation in previous works \citep{Nie_Nature_2022, Sur_npjQM_2023} is a direct consequence of not correctly accounting for the residual resistivity contribution \citep{Wiecki_NatComm_2021} (see open symbols in Fig. 2(a).).

In conclusion, we have demonstrated that the CDW in CsV$_3$Sb$_5$ exhibits a colossal response to $c$-axis stress and strongly compete with SC for the same electronic states. We provide direct thermodynamic evidence that the hydrostatic-pressure dependence of these electronic instabilities originates almost entirely from $c$-axis tuning, as suggested by uniaxial strain experiments \citep{Nie_PRB_2021}. The enhancement of \Tc~and decrease of \Tcdw~under $c-$axis compression is in line with the strong shift of the apical Sb-derived vHS towards the Fermi energy \citep{Consiglio_PRB_2022, Ritz_Arxiv_2023}, highlighting the importance of Sb-derived bands in any minimal microscopic description of this system \citep{Mortensen_PRB_2022, Ritz_Arxiv_2023, Turan_PRB_2023}. Besides CDW and SC, we find no thermodynamic evidence of additional bulk phase transition within the charge-ordered state that could be related to the changes of the $c$-axis periodicity, as reported by x-ray diffraction. The lack of an orthorhombic distortion and the negligibly small $m_{\rm{E_{2g}}}$ response for $T\leq$ \Tcdw~consistently demonstrate (i) the absence of a broken rotational symmetry that couples to anisotropic strain and (ii) that six-fold symmetry is preserved within the individual V$_{3}$Sb$_{5}$ layers. Our results remains, however, consistent with x-ray diffraction \citep{Stahl_PRB_2022} if the reported breaking of C$_6$ invariance arises from a stacking of different CDW patterns along the $c$-axis, as suggested theoretically in Ref. \citep{Ritz_Arxiv_2023}. Our data further show that the large elastoresistance, previously assigned to the nematic $E_{2g}$ channel, definitively originates from an enhanced $A_{1g}$ symmetry-preserving channel, that emerges from electron-electron scattering within the CDW phase. A careful comparison of thermodynamic and spectroscopic experiments under $c-$axis compression is a promising way to shed light on the microscopic origin of the CDW and SC formation in AV$_3$Sb$_5$.

{\bf Note added in proof.} After the completion of the manuscript, we became aware of the preprints of Liu {\it et al.} \citep{liu2023} and Asaba {\it et al.} \citep{asaba2023}. We share the conclusions of Liu {\it et al.} \citep{liu2023} about the absence of nematicity within the CDW state of CsV$_3$Sb$_5$. However, our thermal-expansion results and the elastocaloric measurements of Liu {\it et al.} \citep{liu2023} are at odds concerning the putative crystal-symmetry breaking well above the CDW transition reported by Asaba {\it et al.} \citep{asaba2023}.

We acknowledge fruitful discussions with R. M. Fernandes. Work at KIT was partially funded by the Deutsche Forschungsgemeinschaft (DFG, German Research
Foundation) TRR 288-422213477 (Projects A02 and B03) and Chinesisch-Deutsche Mobilit\"atsprogamm of Chinesisch-Deutsche Zentrum f\"ur Wissenschaftsf\"orderung (Grant No. M-0496). M.F acknowledges funding by the Alexander von Humboldt fundation and the Young Investigator preparation program of the Karlsruhe Institute of Technology. Y. G. acknowledges the support by the National Natural Science Foundation of China (Grant No. 920651). W. X. thanks the support by the Shanghai Sailing Program (23YF1426900). \\

\bibliography{apssamp}

\providecommand{\noopsort}[1]{}\providecommand{\singleletter}[1]{#1}%
\begin{thebibliography}{39}%
\makeatletter
\providecommand \@ifxundefined [1]{%
 \@ifx{#1\undefined}
}%
\providecommand \@ifnum [1]{%
 \ifnum #1\expandafter \@firstoftwo
 \else \expandafter \@secondoftwo
 \fi
}%
\providecommand \@ifx [1]{%
 \ifx #1\expandafter \@firstoftwo
 \else \expandafter \@secondoftwo
 \fi
}%
\providecommand \natexlab [1]{#1}%
\providecommand \enquote  [1]{``#1''}%
\providecommand \bibnamefont  [1]{#1}%
\providecommand \bibfnamefont [1]{#1}%
\providecommand \citenamefont [1]{#1}%
\providecommand \href@noop [0]{\@secondoftwo}%
\providecommand \href [0]{\begingroup \@sanitize@url \@href}%
\providecommand \@href[1]{\@@startlink{#1}\@@href}%
\providecommand \@@href[1]{\endgroup#1\@@endlink}%
\providecommand \@sanitize@url [0]{\catcode `\\12\catcode `\$12\catcode
  `\&12\catcode `\#12\catcode `\^12\catcode `\_12\catcode `\%12\relax}%
\providecommand \@@startlink[1]{}%
\providecommand \@@endlink[0]{}%
\providecommand \url  [0]{\begingroup\@sanitize@url \@url }%
\providecommand \@url [1]{\endgroup\@href {#1}{\urlprefix }}%
\providecommand \urlprefix  [0]{URL }%
\providecommand \Eprint [0]{\href }%
\providecommand \doibase [0]{https://doi.org/}%
\providecommand \selectlanguage [0]{\@gobble}%
\providecommand \bibinfo  [0]{\@secondoftwo}%
\providecommand \bibfield  [0]{\@secondoftwo}%
\providecommand \translation [1]{[#1]}%
\providecommand \BibitemOpen [0]{}%
\providecommand \bibitemStop [0]{}%
\providecommand \bibitemNoStop [0]{.\EOS\space}%
\providecommand \EOS [0]{\spacefactor3000\relax}%
\providecommand \BibitemShut  [1]{\csname bibitem#1\endcsname}%
\let\auto@bib@innerbib\@empty
\bibitem [{\citenamefont {Hu}\ \emph {et~al.}(2022)\citenamefont {Hu},
  \citenamefont {Wu}, \citenamefont {Ortiz}, \citenamefont {Ju}, \citenamefont
  {Han}, \citenamefont {Ma}, \citenamefont {Plumb}, \citenamefont {Radovic},
  \citenamefont {Thomale}, \citenamefont {Wilson}, \citenamefont {Schnyder},\
  and\ \citenamefont {Shi}}]{Hu_NatComm_2022}%
  \BibitemOpen
  \bibfield  {author} {\bibinfo {author} {\bibfnamefont {Y.}~\bibnamefont
  {Hu}}, \bibinfo {author} {\bibfnamefont {X.}~\bibnamefont {Wu}}, \bibinfo
  {author} {\bibfnamefont {B.~R.}\ \bibnamefont {Ortiz}}, \bibinfo {author}
  {\bibfnamefont {S.}~\bibnamefont {Ju}}, \bibinfo {author} {\bibfnamefont
  {X.}~\bibnamefont {Han}}, \bibinfo {author} {\bibfnamefont {J.}~\bibnamefont
  {Ma}}, \bibinfo {author} {\bibfnamefont {N.~C.}\ \bibnamefont {Plumb}},
  \bibinfo {author} {\bibfnamefont {M.}~\bibnamefont {Radovic}}, \bibinfo
  {author} {\bibfnamefont {R.}~\bibnamefont {Thomale}}, \bibinfo {author}
  {\bibfnamefont {S.~D.}\ \bibnamefont {Wilson}}, \bibinfo {author}
  {\bibfnamefont {A.~P.}\ \bibnamefont {Schnyder}},\ and\ \bibinfo {author}
  {\bibfnamefont {M.}~\bibnamefont {Shi}},\ }\bibfield  {title} {\bibinfo
  {title} {Rich nature of van hove singularities in kagome superconductor
  {${\mathrm{CsV}}_{3}{\mathrm{Sb}}_{5}$}},\ }\href
  {https://doi.org/10.1038/s41467-022-29828-x} {\bibfield  {journal} {\bibinfo
  {journal} {Nature Communications}\ }\textbf {\bibinfo {volume} {13}},\
  \bibinfo {pages} {2220} (\bibinfo {year} {2022})}\BibitemShut {NoStop}%
\bibitem [{\citenamefont {Kiesel}\ \emph {et~al.}(2013)\citenamefont {Kiesel},
  \citenamefont {Platt},\ and\ \citenamefont {Thomale}}]{Kiesel_PRL_2013}%
  \BibitemOpen
  \bibfield  {author} {\bibinfo {author} {\bibfnamefont {M.~L.}\ \bibnamefont
  {Kiesel}}, \bibinfo {author} {\bibfnamefont {C.}~\bibnamefont {Platt}},\ and\
  \bibinfo {author} {\bibfnamefont {R.}~\bibnamefont {Thomale}},\ }\bibfield
  {title} {\bibinfo {title} {Unconventional fermi surface instabilities in the
  kagome hubbard model},\ }\href
  {https://doi.org/10.1103/PhysRevLett.110.126405} {\bibfield  {journal}
  {\bibinfo  {journal} {Phys. Rev. Lett.}\ }\textbf {\bibinfo {volume} {110}},\
  \bibinfo {pages} {126405} (\bibinfo {year} {2013})}\BibitemShut {NoStop}%
\bibitem [{\citenamefont {Denner}\ \emph {et~al.}(2021)\citenamefont {Denner},
  \citenamefont {Thomale},\ and\ \citenamefont {Neupert}}]{Denner_PRL_2021}%
  \BibitemOpen
  \bibfield  {author} {\bibinfo {author} {\bibfnamefont {M.~M.}\ \bibnamefont
  {Denner}}, \bibinfo {author} {\bibfnamefont {R.}~\bibnamefont {Thomale}},\
  and\ \bibinfo {author} {\bibfnamefont {T.}~\bibnamefont {Neupert}},\
  }\bibfield  {title} {\bibinfo {title} {Analysis of charge order in the kagome
  metal {$A{\mathrm{V}}_{3}{\mathrm{Sb}}_{5}$}
  ($a=\mathrm{K},\mathrm{Rb},\mathrm{Cs}$)},\ }\href
  {https://doi.org/10.1103/PhysRevLett.127.217601} {\bibfield  {journal}
  {\bibinfo  {journal} {Phys. Rev. Lett.}\ }\textbf {\bibinfo {volume} {127}},\
  \bibinfo {pages} {217601} (\bibinfo {year} {2021})}\BibitemShut {NoStop}%
\bibitem [{\citenamefont {Wang}\ \emph {et~al.}(2013)\citenamefont {Wang},
  \citenamefont {Li}, \citenamefont {Xiang},\ and\ \citenamefont
  {Wang}}]{Shen_PRB_2013}%
  \BibitemOpen
  \bibfield  {author} {\bibinfo {author} {\bibfnamefont {W.-S.}\ \bibnamefont
  {Wang}}, \bibinfo {author} {\bibfnamefont {Z.-Z.}\ \bibnamefont {Li}},
  \bibinfo {author} {\bibfnamefont {Y.-Y.}\ \bibnamefont {Xiang}},\ and\
  \bibinfo {author} {\bibfnamefont {Q.-H.}\ \bibnamefont {Wang}},\ }\bibfield
  {title} {\bibinfo {title} {Competing electronic orders on kagome lattices at
  van hove filling},\ }\href {https://doi.org/10.1103/PhysRevB.87.115135}
  {\bibfield  {journal} {\bibinfo  {journal} {Phys. Rev. B}\ }\textbf {\bibinfo
  {volume} {87}},\ \bibinfo {pages} {115135} (\bibinfo {year}
  {2013})}\BibitemShut {NoStop}%
\bibitem [{\citenamefont {Ortiz}\ \emph {et~al.}(2019)\citenamefont {Ortiz},
  \citenamefont {Gomes}, \citenamefont {Morey}, \citenamefont {Winiarski},
  \citenamefont {Bordelon}, \citenamefont {Mangum}, \citenamefont {Oswald},
  \citenamefont {Rodriguez-Rivera}, \citenamefont {Neilson}, \citenamefont
  {Wilson}, \citenamefont {Ertekin}, \citenamefont {McQueen},\ and\
  \citenamefont {Toberer}}]{Ortiz_PRM_2019}%
  \BibitemOpen
  \bibfield  {author} {\bibinfo {author} {\bibfnamefont {B.~R.}\ \bibnamefont
  {Ortiz}}, \bibinfo {author} {\bibfnamefont {L.~C.}\ \bibnamefont {Gomes}},
  \bibinfo {author} {\bibfnamefont {J.~R.}\ \bibnamefont {Morey}}, \bibinfo
  {author} {\bibfnamefont {M.}~\bibnamefont {Winiarski}}, \bibinfo {author}
  {\bibfnamefont {M.}~\bibnamefont {Bordelon}}, \bibinfo {author}
  {\bibfnamefont {J.~S.}\ \bibnamefont {Mangum}}, \bibinfo {author}
  {\bibfnamefont {I.~W.~H.}\ \bibnamefont {Oswald}}, \bibinfo {author}
  {\bibfnamefont {J.~A.}\ \bibnamefont {Rodriguez-Rivera}}, \bibinfo {author}
  {\bibfnamefont {J.~R.}\ \bibnamefont {Neilson}}, \bibinfo {author}
  {\bibfnamefont {S.~D.}\ \bibnamefont {Wilson}}, \bibinfo {author}
  {\bibfnamefont {E.}~\bibnamefont {Ertekin}}, \bibinfo {author} {\bibfnamefont
  {T.~M.}\ \bibnamefont {McQueen}},\ and\ \bibinfo {author} {\bibfnamefont
  {E.~S.}\ \bibnamefont {Toberer}},\ }\bibfield  {title} {\bibinfo {title} {New
  kagome prototype materials: discovery of
  {${\mathrm{KV}}_{3}{\mathrm{Sb}}_{5}$},{${\mathrm{RbV}}_{3}{\mathrm{Sb}}_{5}$},
  and {${\mathrm{CsV}}_{3}{\mathrm{Sb}}_{5}$}},\ }\href
  {https://doi.org/10.1103/PhysRevMaterials.3.094407} {\bibfield  {journal}
  {\bibinfo  {journal} {Phys. Rev. Mater.}\ }\textbf {\bibinfo {volume} {3}},\
  \bibinfo {pages} {094407} (\bibinfo {year} {2019})}\BibitemShut {NoStop}%
\bibitem [{\citenamefont {Ortiz}\ \emph {et~al.}(2020)\citenamefont {Ortiz},
  \citenamefont {Teicher}, \citenamefont {Hu}, \citenamefont {Zuo},
  \citenamefont {Sarte}, \citenamefont {Schueller}, \citenamefont {Abeykoon},
  \citenamefont {Krogstad}, \citenamefont {Rosenkranz}, \citenamefont {Osborn},
  \citenamefont {Seshadri}, \citenamefont {Balents}, \citenamefont {He},\ and\
  \citenamefont {Wilson}}]{Ortiz_PRL_2020}%
  \BibitemOpen
  \bibfield  {author} {\bibinfo {author} {\bibfnamefont {B.~R.}\ \bibnamefont
  {Ortiz}}, \bibinfo {author} {\bibfnamefont {S.~M.~L.}\ \bibnamefont
  {Teicher}}, \bibinfo {author} {\bibfnamefont {Y.}~\bibnamefont {Hu}},
  \bibinfo {author} {\bibfnamefont {J.~L.}\ \bibnamefont {Zuo}}, \bibinfo
  {author} {\bibfnamefont {P.~M.}\ \bibnamefont {Sarte}}, \bibinfo {author}
  {\bibfnamefont {E.~C.}\ \bibnamefont {Schueller}}, \bibinfo {author}
  {\bibfnamefont {A.~M.~M.}\ \bibnamefont {Abeykoon}}, \bibinfo {author}
  {\bibfnamefont {M.~J.}\ \bibnamefont {Krogstad}}, \bibinfo {author}
  {\bibfnamefont {S.}~\bibnamefont {Rosenkranz}}, \bibinfo {author}
  {\bibfnamefont {R.}~\bibnamefont {Osborn}}, \bibinfo {author} {\bibfnamefont
  {R.}~\bibnamefont {Seshadri}}, \bibinfo {author} {\bibfnamefont
  {L.}~\bibnamefont {Balents}}, \bibinfo {author} {\bibfnamefont
  {J.}~\bibnamefont {He}},\ and\ \bibinfo {author} {\bibfnamefont {S.~D.}\
  \bibnamefont {Wilson}},\ }\bibfield  {title} {\bibinfo {title}
  {{$\mathrm{Cs}{\mathrm{V}}_{3}{\mathrm{Sb}}_{5}$}: A {${\mathbb{Z}}_{2}$}
  topological kagome metal with a superconducting ground state},\ }\href
  {https://doi.org/10.1103/PhysRevLett.125.247002} {\bibfield  {journal}
  {\bibinfo  {journal} {Phys. Rev. Lett.}\ }\textbf {\bibinfo {volume} {125}},\
  \bibinfo {pages} {247002} (\bibinfo {year} {2020})}\BibitemShut {NoStop}%
\bibitem [{\citenamefont {Liang}\ \emph {et~al.}(2021)\citenamefont {Liang},
  \citenamefont {Hou}, \citenamefont {Zhang}, \citenamefont {Ma}, \citenamefont
  {Wu}, \citenamefont {Zhang}, \citenamefont {Yu}, \citenamefont {Ying},
  \citenamefont {Jiang}, \citenamefont {Shan}, \citenamefont {Wang},\ and\
  \citenamefont {Chen}}]{Liang_PRX_2021}%
  \BibitemOpen
  \bibfield  {author} {\bibinfo {author} {\bibfnamefont {Z.}~\bibnamefont
  {Liang}}, \bibinfo {author} {\bibfnamefont {X.}~\bibnamefont {Hou}}, \bibinfo
  {author} {\bibfnamefont {F.}~\bibnamefont {Zhang}}, \bibinfo {author}
  {\bibfnamefont {W.}~\bibnamefont {Ma}}, \bibinfo {author} {\bibfnamefont
  {P.}~\bibnamefont {Wu}}, \bibinfo {author} {\bibfnamefont {Z.}~\bibnamefont
  {Zhang}}, \bibinfo {author} {\bibfnamefont {F.}~\bibnamefont {Yu}}, \bibinfo
  {author} {\bibfnamefont {J.-J.}\ \bibnamefont {Ying}}, \bibinfo {author}
  {\bibfnamefont {K.}~\bibnamefont {Jiang}}, \bibinfo {author} {\bibfnamefont
  {L.}~\bibnamefont {Shan}}, \bibinfo {author} {\bibfnamefont {Z.}~\bibnamefont
  {Wang}},\ and\ \bibinfo {author} {\bibfnamefont {X.-H.}\ \bibnamefont
  {Chen}},\ }\bibfield  {title} {\bibinfo {title} {Three-dimensional charge
  density wave and surface-dependent vortex-core states in a kagome
  superconductor ${\mathrm{csv}}_{3}{\mathrm{sb}}_{5}$},\ }\href
  {https://doi.org/10.1103/PhysRevX.11.031026} {\bibfield  {journal} {\bibinfo
  {journal} {Phys. Rev. X}\ }\textbf {\bibinfo {volume} {11}},\ \bibinfo
  {pages} {031026} (\bibinfo {year} {2021})}\BibitemShut {NoStop}%
\bibitem [{\citenamefont {Tan}\ \emph {et~al.}(2021)\citenamefont {Tan},
  \citenamefont {Liu}, \citenamefont {Wang},\ and\ \citenamefont
  {Yan}}]{Tan_PRL_2021}%
  \BibitemOpen
  \bibfield  {author} {\bibinfo {author} {\bibfnamefont {H.}~\bibnamefont
  {Tan}}, \bibinfo {author} {\bibfnamefont {Y.}~\bibnamefont {Liu}}, \bibinfo
  {author} {\bibfnamefont {Z.}~\bibnamefont {Wang}},\ and\ \bibinfo {author}
  {\bibfnamefont {B.}~\bibnamefont {Yan}},\ }\bibfield  {title} {\bibinfo
  {title} {Charge density waves and electronic properties of superconducting
  kagome metals},\ }\href {https://doi.org/10.1103/PhysRevLett.127.046401}
  {\bibfield  {journal} {\bibinfo  {journal} {Phys. Rev. Lett.}\ }\textbf
  {\bibinfo {volume} {127}},\ \bibinfo {pages} {046401} (\bibinfo {year}
  {2021})}\BibitemShut {NoStop}%
\bibitem [{\citenamefont {Ortiz}\ \emph {et~al.}(2021)\citenamefont {Ortiz},
  \citenamefont {Teicher}, \citenamefont {Kautzsch}, \citenamefont {Sarte},
  \citenamefont {Ratcliff}, \citenamefont {Harter}, \citenamefont {Ruff},
  \citenamefont {Seshadri},\ and\ \citenamefont {Wilson}}]{Ortiz_PRX_2021}%
  \BibitemOpen
  \bibfield  {author} {\bibinfo {author} {\bibfnamefont {B.~R.}\ \bibnamefont
  {Ortiz}}, \bibinfo {author} {\bibfnamefont {S.~M.~L.}\ \bibnamefont
  {Teicher}}, \bibinfo {author} {\bibfnamefont {L.}~\bibnamefont {Kautzsch}},
  \bibinfo {author} {\bibfnamefont {P.~M.}\ \bibnamefont {Sarte}}, \bibinfo
  {author} {\bibfnamefont {N.}~\bibnamefont {Ratcliff}}, \bibinfo {author}
  {\bibfnamefont {J.}~\bibnamefont {Harter}}, \bibinfo {author} {\bibfnamefont
  {J.~P.~C.}\ \bibnamefont {Ruff}}, \bibinfo {author} {\bibfnamefont
  {R.}~\bibnamefont {Seshadri}},\ and\ \bibinfo {author} {\bibfnamefont
  {S.~D.}\ \bibnamefont {Wilson}},\ }\bibfield  {title} {\bibinfo {title}
  {Fermi surface mapping and the nature of charge-density-wave order in the
  kagome superconductor ${\mathrm{csv}}_{3}{\mathrm{sb}}_{5}$},\ }\href
  {https://doi.org/10.1103/PhysRevX.11.041030} {\bibfield  {journal} {\bibinfo
  {journal} {Phys. Rev. X}\ }\textbf {\bibinfo {volume} {11}},\ \bibinfo
  {pages} {041030} (\bibinfo {year} {2021})}\BibitemShut {NoStop}%
\bibitem [{\citenamefont {Xiao}\ \emph {et~al.}(2023)\citenamefont {Xiao},
  \citenamefont {Lin}, \citenamefont {Li}, \citenamefont {Zheng}, \citenamefont
  {Francoual}, \citenamefont {Plueckthun}, \citenamefont {Xia}, \citenamefont
  {Qiu}, \citenamefont {Zhang}, \citenamefont {Guo}, \citenamefont {Feng},\
  and\ \citenamefont {Peng}}]{Xiao_PRX_2023}%
  \BibitemOpen
  \bibfield  {author} {\bibinfo {author} {\bibfnamefont {Q.}~\bibnamefont
  {Xiao}}, \bibinfo {author} {\bibfnamefont {Y.}~\bibnamefont {Lin}}, \bibinfo
  {author} {\bibfnamefont {Q.}~\bibnamefont {Li}}, \bibinfo {author}
  {\bibfnamefont {X.}~\bibnamefont {Zheng}}, \bibinfo {author} {\bibfnamefont
  {S.}~\bibnamefont {Francoual}}, \bibinfo {author} {\bibfnamefont
  {C.}~\bibnamefont {Plueckthun}}, \bibinfo {author} {\bibfnamefont
  {W.}~\bibnamefont {Xia}}, \bibinfo {author} {\bibfnamefont {Q.}~\bibnamefont
  {Qiu}}, \bibinfo {author} {\bibfnamefont {S.}~\bibnamefont {Zhang}}, \bibinfo
  {author} {\bibfnamefont {Y.}~\bibnamefont {Guo}}, \bibinfo {author}
  {\bibfnamefont {J.}~\bibnamefont {Feng}},\ and\ \bibinfo {author}
  {\bibfnamefont {Y.}~\bibnamefont {Peng}},\ }\bibfield  {title} {\bibinfo
  {title} {Coexistence of multiple stacking charge density waves in kagome
  superconductor {${\mathrm{CsV}}_{3}{\mathrm{Sb}}_{5}$}},\ }\href
  {https://doi.org/10.1103/PhysRevResearch.5.L012032} {\bibfield  {journal}
  {\bibinfo  {journal} {Phys. Rev. Res.}\ }\textbf {\bibinfo {volume} {5}},\
  \bibinfo {pages} {L012032} (\bibinfo {year} {2023})}\BibitemShut {NoStop}%
\bibitem [{\citenamefont {Stahl}\ \emph {et~al.}(2022)\citenamefont {Stahl},
  \citenamefont {Chen}, \citenamefont {Ritschel}, \citenamefont {Shekhar},
  \citenamefont {Sadrollahi}, \citenamefont {Rahn}, \citenamefont {Ivashko},
  \citenamefont {Zimmermann}, \citenamefont {Felser},\ and\ \citenamefont
  {Geck}}]{Stahl_PRB_2022}%
  \BibitemOpen
  \bibfield  {author} {\bibinfo {author} {\bibfnamefont {Q.}~\bibnamefont
  {Stahl}}, \bibinfo {author} {\bibfnamefont {D.}~\bibnamefont {Chen}},
  \bibinfo {author} {\bibfnamefont {T.}~\bibnamefont {Ritschel}}, \bibinfo
  {author} {\bibfnamefont {C.}~\bibnamefont {Shekhar}}, \bibinfo {author}
  {\bibfnamefont {E.}~\bibnamefont {Sadrollahi}}, \bibinfo {author}
  {\bibfnamefont {M.~C.}\ \bibnamefont {Rahn}}, \bibinfo {author}
  {\bibfnamefont {O.}~\bibnamefont {Ivashko}}, \bibinfo {author} {\bibfnamefont
  {M.~v.}\ \bibnamefont {Zimmermann}}, \bibinfo {author} {\bibfnamefont
  {C.}~\bibnamefont {Felser}},\ and\ \bibinfo {author} {\bibfnamefont
  {J.}~\bibnamefont {Geck}},\ }\bibfield  {title} {\bibinfo {title}
  {Temperature-driven reorganization of electronic order in
  {${\mathrm{CsV}}_{3}{\mathrm{Sb}}_{5}$}},\ }\href
  {https://doi.org/10.1103/PhysRevB.105.195136} {\bibfield  {journal} {\bibinfo
   {journal} {Phys. Rev. B}\ }\textbf {\bibinfo {volume} {105}},\ \bibinfo
  {pages} {195136} (\bibinfo {year} {2022})}\BibitemShut {NoStop}%
\bibitem [{\citenamefont {Xu}\ \emph {et~al.}(2022{\natexlab{a}})\citenamefont
  {Xu}, \citenamefont {Ni}, \citenamefont {Liu}, \citenamefont {Ortiz},
  \citenamefont {Deng}, \citenamefont {Wilson}, \citenamefont {Yan},
  \citenamefont {Balents},\ and\ \citenamefont {Wu}}]{Xu_NatPhys_2022}%
  \BibitemOpen
  \bibfield  {author} {\bibinfo {author} {\bibfnamefont {Y.}~\bibnamefont
  {Xu}}, \bibinfo {author} {\bibfnamefont {Z.}~\bibnamefont {Ni}}, \bibinfo
  {author} {\bibfnamefont {Y.}~\bibnamefont {Liu}}, \bibinfo {author}
  {\bibfnamefont {B.~R.}\ \bibnamefont {Ortiz}}, \bibinfo {author}
  {\bibfnamefont {Q.}~\bibnamefont {Deng}}, \bibinfo {author} {\bibfnamefont
  {S.~D.}\ \bibnamefont {Wilson}}, \bibinfo {author} {\bibfnamefont
  {B.}~\bibnamefont {Yan}}, \bibinfo {author} {\bibfnamefont {L.}~\bibnamefont
  {Balents}},\ and\ \bibinfo {author} {\bibfnamefont {L.}~\bibnamefont {Wu}},\
  }\bibfield  {title} {\bibinfo {title} {Three-state nematicity and
  magneto-optical kerr effect in the charge density waves in kagome
  superconductors},\ }\href {https://doi.org/10.1038/s41567-022-01805-7}
  {\bibfield  {journal} {\bibinfo  {journal} {Nature Physics}\ }\textbf
  {\bibinfo {volume} {18}},\ \bibinfo {pages} {1470} (\bibinfo {year}
  {2022}{\natexlab{a}})}\BibitemShut {NoStop}%
\bibitem [{\citenamefont {Xiang}\ \emph {et~al.}(2021)\citenamefont {Xiang},
  \citenamefont {Li}, \citenamefont {Li}, \citenamefont {Xie}, \citenamefont
  {Yang}, \citenamefont {Wang}, \citenamefont {Yao},\ and\ \citenamefont
  {Wen}}]{Xiang_NatComm_2021}%
  \BibitemOpen
  \bibfield  {author} {\bibinfo {author} {\bibfnamefont {Y.}~\bibnamefont
  {Xiang}}, \bibinfo {author} {\bibfnamefont {Q.}~\bibnamefont {Li}}, \bibinfo
  {author} {\bibfnamefont {Y.}~\bibnamefont {Li}}, \bibinfo {author}
  {\bibfnamefont {W.}~\bibnamefont {Xie}}, \bibinfo {author} {\bibfnamefont
  {H.}~\bibnamefont {Yang}}, \bibinfo {author} {\bibfnamefont {Z.}~\bibnamefont
  {Wang}}, \bibinfo {author} {\bibfnamefont {Y.}~\bibnamefont {Yao}},\ and\
  \bibinfo {author} {\bibfnamefont {H.-H.}\ \bibnamefont {Wen}},\ }\bibfield
  {title} {\bibinfo {title} {Twofold symmetry of c-axis resistivity in
  topological kagome superconductor {${\mathrm{CsV}}_{3}{\mathrm{Sb}}_{5}$}
  with in-plane rotating magnetic field},\ }\href
  {https://doi.org/10.1038/s41467-021-27084-z} {\bibfield  {journal} {\bibinfo
  {journal} {Nature Communications}\ }\textbf {\bibinfo {volume} {12}},\
  \bibinfo {pages} {6727} (\bibinfo {year} {2021})}\BibitemShut {NoStop}%
\bibitem [{\citenamefont {Nie}\ \emph {et~al.}(2022)\citenamefont {Nie},
  \citenamefont {Sun}, \citenamefont {Ma}, \citenamefont {Song}, \citenamefont
  {Zheng}, \citenamefont {Liang}, \citenamefont {Wu}, \citenamefont {Yu},
  \citenamefont {Li}, \citenamefont {Shan}, \citenamefont {Zhao}, \citenamefont
  {Li}, \citenamefont {Kang}, \citenamefont {Wu}, \citenamefont {Y.},
  \citenamefont {Liu}, \citenamefont {Xiang}, \citenamefont {Ying},
  \citenamefont {Wang}, \citenamefont {Wu},\ and\ \citenamefont
  {Chen}}]{Nie_Nature_2022}%
  \BibitemOpen
  \bibfield  {author} {\bibinfo {author} {\bibfnamefont {L.}~\bibnamefont
  {Nie}}, \bibinfo {author} {\bibfnamefont {K.}~\bibnamefont {Sun}}, \bibinfo
  {author} {\bibfnamefont {W.}~\bibnamefont {Ma}}, \bibinfo {author}
  {\bibfnamefont {D.}~\bibnamefont {Song}}, \bibinfo {author} {\bibfnamefont
  {L.}~\bibnamefont {Zheng}}, \bibinfo {author} {\bibfnamefont
  {Z.}~\bibnamefont {Liang}}, \bibinfo {author} {\bibfnamefont
  {P.}~\bibnamefont {Wu}}, \bibinfo {author} {\bibfnamefont {F.}~\bibnamefont
  {Yu}}, \bibinfo {author} {\bibfnamefont {J.}~\bibnamefont {Li}}, \bibinfo
  {author} {\bibfnamefont {M.}~\bibnamefont {Shan}}, \bibinfo {author}
  {\bibfnamefont {D.}~\bibnamefont {Zhao}}, \bibinfo {author} {\bibfnamefont
  {S.}~\bibnamefont {Li}}, \bibinfo {author} {\bibfnamefont {B.}~\bibnamefont
  {Kang}}, \bibinfo {author} {\bibfnamefont {Z.}~\bibnamefont {Wu}}, \bibinfo
  {author} {\bibfnamefont {Z.}~\bibnamefont {Y.}}, \bibinfo {author}
  {\bibfnamefont {K.}~\bibnamefont {Liu}}, \bibinfo {author} {\bibfnamefont
  {Z.}~\bibnamefont {Xiang}}, \bibinfo {author} {\bibfnamefont
  {J.}~\bibnamefont {Ying}}, \bibinfo {author} {\bibfnamefont {Z.}~\bibnamefont
  {Wang}}, \bibinfo {author} {\bibfnamefont {T.}~\bibnamefont {Wu}},\ and\
  \bibinfo {author} {\bibfnamefont {X.}~\bibnamefont {Chen}},\ }\bibfield
  {title} {\bibinfo {title} {Charge-density-wave-driven electronic nematicity
  in a kagome superconductor},\ }\href
  {https://doi.org/10.1038/s41586-022-04493-8} {\bibfield  {journal} {\bibinfo
  {journal} {Nature}\ }\textbf {\bibinfo {volume} {604}},\ \bibinfo {pages}
  {59} (\bibinfo {year} {2022})}\BibitemShut {NoStop}%
\bibitem [{\citenamefont {Sur}\ \emph {et~al.}(2023)\citenamefont {Sur},
  \citenamefont {Kim}, \citenamefont {Kim},\ and\ \citenamefont
  {H.}}]{Sur_npjQM_2023}%
  \BibitemOpen
  \bibfield  {author} {\bibinfo {author} {\bibfnamefont {Y.}~\bibnamefont
  {Sur}}, \bibinfo {author} {\bibfnamefont {K.}~\bibnamefont {Kim}}, \bibinfo
  {author} {\bibfnamefont {S.}~\bibnamefont {Kim}},\ and\ \bibinfo {author}
  {\bibfnamefont {K.~K.}\ \bibnamefont {H.}},\ }\bibfield  {title} {\bibinfo
  {title} {Optimized superconductivity in the vicinity of a nematic quantum
  critical point in the kagome superconductor
  {$\mathrm{Cs}(\mathrm{V}_{1-x}\mathrm{Ti}_{x})_{3}\mathrm{Sb}_{5}$}},\ }\href
  {https://doi.org/10.1038/s41467-023-39495-1} {\bibfield  {journal} {\bibinfo
  {journal} {Nature Communications}\ }\textbf {\bibinfo {volume} {14}},\
  \bibinfo {pages} {3899} (\bibinfo {year} {2023})}\BibitemShut {NoStop}%
\bibitem [{\citenamefont {Saykin}\ \emph {et~al.}(2023)\citenamefont {Saykin},
  \citenamefont {Farhang}, \citenamefont {Kountz}, \citenamefont {Chen},
  \citenamefont {Ortiz}, \citenamefont {Shekhar}, \citenamefont {Felser},
  \citenamefont {Wilson}, \citenamefont {Thomale}, \citenamefont {Xia},\ and\
  \citenamefont {Kapitulnik}}]{Saykin_PRL_2023}%
  \BibitemOpen
  \bibfield  {author} {\bibinfo {author} {\bibfnamefont {D.~R.}\ \bibnamefont
  {Saykin}}, \bibinfo {author} {\bibfnamefont {C.}~\bibnamefont {Farhang}},
  \bibinfo {author} {\bibfnamefont {E.~D.}\ \bibnamefont {Kountz}}, \bibinfo
  {author} {\bibfnamefont {D.}~\bibnamefont {Chen}}, \bibinfo {author}
  {\bibfnamefont {B.~R.}\ \bibnamefont {Ortiz}}, \bibinfo {author}
  {\bibfnamefont {C.}~\bibnamefont {Shekhar}}, \bibinfo {author} {\bibfnamefont
  {C.}~\bibnamefont {Felser}}, \bibinfo {author} {\bibfnamefont {S.~D.}\
  \bibnamefont {Wilson}}, \bibinfo {author} {\bibfnamefont {R.}~\bibnamefont
  {Thomale}}, \bibinfo {author} {\bibfnamefont {J.}~\bibnamefont {Xia}},\ and\
  \bibinfo {author} {\bibfnamefont {A.}~\bibnamefont {Kapitulnik}},\ }\bibfield
   {title} {\bibinfo {title} {High resolution polar kerr effect studies of
  ${\mathrm{csv}}_{3}{\mathrm{sb}}_{5}$: Tests for time-reversal symmetry
  breaking below the charge-order transition},\ }\href
  {https://doi.org/10.1103/PhysRevLett.131.016901} {\bibfield  {journal}
  {\bibinfo  {journal} {Phys. Rev. Lett.}\ }\textbf {\bibinfo {volume} {131}},\
  \bibinfo {pages} {016901} (\bibinfo {year} {2023})}\BibitemShut {NoStop}%
\bibitem [{\citenamefont {Hu}\ \emph {et~al.}(2023)\citenamefont {Hu},
  \citenamefont {Yamane}, \citenamefont {Mattoni}, \citenamefont {Yada},
  \citenamefont {Obata}, \citenamefont {Li}, \citenamefont {Yao}, \citenamefont
  {Wang}, \citenamefont {Wang}, \citenamefont {Farhang}, \citenamefont {Xia},
  \citenamefont {Maeno},\ and\ \citenamefont {Yonezawa}}]{Hu_Arxiv_2023}%
  \BibitemOpen
  \bibfield  {author} {\bibinfo {author} {\bibfnamefont {Y.}~\bibnamefont
  {Hu}}, \bibinfo {author} {\bibfnamefont {S.}~\bibnamefont {Yamane}}, \bibinfo
  {author} {\bibfnamefont {G.}~\bibnamefont {Mattoni}}, \bibinfo {author}
  {\bibfnamefont {K.}~\bibnamefont {Yada}}, \bibinfo {author} {\bibfnamefont
  {K.}~\bibnamefont {Obata}}, \bibinfo {author} {\bibfnamefont
  {Y.}~\bibnamefont {Li}}, \bibinfo {author} {\bibfnamefont {Y.}~\bibnamefont
  {Yao}}, \bibinfo {author} {\bibfnamefont {Z.}~\bibnamefont {Wang}}, \bibinfo
  {author} {\bibfnamefont {J.}~\bibnamefont {Wang}}, \bibinfo {author}
  {\bibfnamefont {C.}~\bibnamefont {Farhang}}, \bibinfo {author} {\bibfnamefont
  {J.}~\bibnamefont {Xia}}, \bibinfo {author} {\bibfnamefont {Y.}~\bibnamefont
  {Maeno}},\ and\ \bibinfo {author} {\bibfnamefont {S.}~\bibnamefont
  {Yonezawa}},\ }\href@noop {} {\bibinfo {title} {Time-reversal symmetry
  breaking in charge density wave of {${\mathrm{CsV}}_{3}{\mathrm{Sb}}_{5}$}
  detected by polar kerr effect}} (\bibinfo {year} {2023}),\ \Eprint
  {https://arxiv.org/abs/2208.08036} {arXiv:2208.08036 [cond-mat.str-el]}
  \BibitemShut {NoStop}%
\bibitem [{\citenamefont {Xu}\ \emph {et~al.}(2022{\natexlab{b}})\citenamefont
  {Xu}, \citenamefont {Ni}, \citenamefont {Liu}, \citenamefont {Ortiz},
  \citenamefont {Deng}, \citenamefont {Wilson}, \citenamefont {Yan},
  \citenamefont {Balents},\ and\ \citenamefont {Wu}}]{Xu_Nature_2022}%
  \BibitemOpen
  \bibfield  {author} {\bibinfo {author} {\bibfnamefont {Y.}~\bibnamefont
  {Xu}}, \bibinfo {author} {\bibfnamefont {Z.}~\bibnamefont {Ni}}, \bibinfo
  {author} {\bibfnamefont {Y.}~\bibnamefont {Liu}}, \bibinfo {author}
  {\bibfnamefont {B.~R.}\ \bibnamefont {Ortiz}}, \bibinfo {author}
  {\bibfnamefont {Q.}~\bibnamefont {Deng}}, \bibinfo {author} {\bibfnamefont
  {S.~D.}\ \bibnamefont {Wilson}}, \bibinfo {author} {\bibfnamefont
  {B.}~\bibnamefont {Yan}}, \bibinfo {author} {\bibfnamefont {L.}~\bibnamefont
  {Balents}},\ and\ \bibinfo {author} {\bibfnamefont {L.}~\bibnamefont {Wu}},\
  }\bibfield  {title} {\bibinfo {title} {Three-state nematicity and
  magneto-optical kerr effect in the charge density waves in kagome
  superconductors},\ }\href {https://doi.org/10.1038/s41567-022-01805-7}
  {\bibfield  {journal} {\bibinfo  {journal} {Nature}\ }\textbf {\bibinfo
  {volume} {18}},\ \bibinfo {pages} {1470} (\bibinfo {year}
  {2022}{\natexlab{b}})}\BibitemShut {NoStop}%
\bibitem [{\citenamefont {Mielke}\ \emph {et~al.}(2022)\citenamefont {Mielke},
  \citenamefont {Das}, \citenamefont {Yin}, \citenamefont {Liu}, \citenamefont
  {Gupta}, \citenamefont {Jiang}, \citenamefont {Medarde}, \citenamefont {Wu},
  \citenamefont {Lei}, \citenamefont {Chang}, \citenamefont {Dai},
  \citenamefont {Si}, \citenamefont {Miao}, \citenamefont {Thomale},
  \citenamefont {Neupert}, \citenamefont {Shi}, \citenamefont {Khasanov},
  \citenamefont {Hasan}, \citenamefont {Luetkens},\ and\ \citenamefont
  {Guguchia}}]{Mielke_Nature_2022}%
  \BibitemOpen
  \bibfield  {author} {\bibinfo {author} {\bibfnamefont {C.}~\bibnamefont
  {Mielke}}, \bibinfo {author} {\bibfnamefont {D.}~\bibnamefont {Das}},
  \bibinfo {author} {\bibfnamefont {J.-X.}\ \bibnamefont {Yin}}, \bibinfo
  {author} {\bibfnamefont {H.}~\bibnamefont {Liu}}, \bibinfo {author}
  {\bibfnamefont {R.}~\bibnamefont {Gupta}}, \bibinfo {author} {\bibfnamefont
  {Y.-X.}\ \bibnamefont {Jiang}}, \bibinfo {author} {\bibfnamefont
  {M.}~\bibnamefont {Medarde}}, \bibinfo {author} {\bibfnamefont
  {X.}~\bibnamefont {Wu}}, \bibinfo {author} {\bibfnamefont {H.~C.}\
  \bibnamefont {Lei}}, \bibinfo {author} {\bibfnamefont {J.}~\bibnamefont
  {Chang}}, \bibinfo {author} {\bibfnamefont {P.}~\bibnamefont {Dai}}, \bibinfo
  {author} {\bibfnamefont {Q.}~\bibnamefont {Si}}, \bibinfo {author}
  {\bibfnamefont {H.}~\bibnamefont {Miao}}, \bibinfo {author} {\bibfnamefont
  {R.}~\bibnamefont {Thomale}}, \bibinfo {author} {\bibfnamefont
  {T.}~\bibnamefont {Neupert}}, \bibinfo {author} {\bibfnamefont
  {Y.}~\bibnamefont {Shi}}, \bibinfo {author} {\bibfnamefont {R.}~\bibnamefont
  {Khasanov}}, \bibinfo {author} {\bibfnamefont {M.~Z.}\ \bibnamefont {Hasan}},
  \bibinfo {author} {\bibfnamefont {H.}~\bibnamefont {Luetkens}},\ and\
  \bibinfo {author} {\bibfnamefont {Z.}~\bibnamefont {Guguchia}},\ }\bibfield
  {title} {\bibinfo {title} {Time-reversal symmetry-breaking charge order in a
  kagome superconductor},\ }\href {https://doi.org/10.1038/s41586-021-04327-z}
  {\bibfield  {journal} {\bibinfo  {journal} {Nature}\ }\textbf {\bibinfo
  {volume} {602}},\ \bibinfo {pages} {245} (\bibinfo {year}
  {2022})}\BibitemShut {NoStop}%
\bibitem [{\citenamefont {Christensen}\ \emph {et~al.}(2022)\citenamefont
  {Christensen}, \citenamefont {Birol}, \citenamefont {Andersen},\ and\
  \citenamefont {Fernandes}}]{Mortensen_PRB_2022}%
  \BibitemOpen
  \bibfield  {author} {\bibinfo {author} {\bibfnamefont {M.~H.}\ \bibnamefont
  {Christensen}}, \bibinfo {author} {\bibfnamefont {T.}~\bibnamefont {Birol}},
  \bibinfo {author} {\bibfnamefont {B.~M.}\ \bibnamefont {Andersen}},\ and\
  \bibinfo {author} {\bibfnamefont {R.~M.}\ \bibnamefont {Fernandes}},\
  }\bibfield  {title} {\bibinfo {title} {Loop currents in
  {$A{\mathrm{V}}_{3}{\mathrm{Sb}}_{5}$} kagome metals: Multipolar and toroidal
  magnetic orders},\ }\href {https://doi.org/10.1103/PhysRevB.106.144504}
  {\bibfield  {journal} {\bibinfo  {journal} {Phys. Rev. B}\ }\textbf {\bibinfo
  {volume} {106}},\ \bibinfo {pages} {144504} (\bibinfo {year}
  {2022})}\BibitemShut {NoStop}%
\bibitem [{\citenamefont {Zhao}\ \emph {et~al.}(2021)\citenamefont {Zhao},
  \citenamefont {Wang}, \citenamefont {Xia}, \citenamefont {Yin}, \citenamefont
  {Ni}, \citenamefont {Huang}, \citenamefont {Tu}, \citenamefont {Tao},
  \citenamefont {Tu}, \citenamefont {Gong}, \citenamefont {Lei}, \citenamefont
  {Guo}, \citenamefont {Yang},\ and\ \citenamefont {Li}}]{Zhao_Arxiv_2021}%
  \BibitemOpen
  \bibfield  {author} {\bibinfo {author} {\bibfnamefont {C.~C.}\ \bibnamefont
  {Zhao}}, \bibinfo {author} {\bibfnamefont {L.~S.}\ \bibnamefont {Wang}},
  \bibinfo {author} {\bibfnamefont {W.}~\bibnamefont {Xia}}, \bibinfo {author}
  {\bibfnamefont {Q.~W.}\ \bibnamefont {Yin}}, \bibinfo {author} {\bibfnamefont
  {J.~M.}\ \bibnamefont {Ni}}, \bibinfo {author} {\bibfnamefont {Y.~Y.}\
  \bibnamefont {Huang}}, \bibinfo {author} {\bibfnamefont {C.~P.}\ \bibnamefont
  {Tu}}, \bibinfo {author} {\bibfnamefont {Z.~C.}\ \bibnamefont {Tao}},
  \bibinfo {author} {\bibfnamefont {Z.~J.}\ \bibnamefont {Tu}}, \bibinfo
  {author} {\bibfnamefont {C.~S.}\ \bibnamefont {Gong}}, \bibinfo {author}
  {\bibfnamefont {H.~C.}\ \bibnamefont {Lei}}, \bibinfo {author} {\bibfnamefont
  {Y.~F.}\ \bibnamefont {Guo}}, \bibinfo {author} {\bibfnamefont {X.~F.}\
  \bibnamefont {Yang}},\ and\ \bibinfo {author} {\bibfnamefont {S.~Y.}\
  \bibnamefont {Li}},\ }\href@noop {} {\bibinfo {title} {Nodal
  superconductivity and superconducting domes in the topological kagome metal
  {${\mathrm{CsV}}_{3}{\mathrm{Sb}}_{5}$}}} (\bibinfo {year} {2021}),\ \Eprint
  {https://arxiv.org/abs/2102.08356} {arXiv:2102.08356 [cond-mat.supr-con]}
  \BibitemShut {NoStop}%
\bibitem [{\citenamefont {Duan}\ \emph {et~al.}(2021)\citenamefont {Duan},
  \citenamefont {Nie}, \citenamefont {Luo}, \citenamefont {Yu}, \citenamefont
  {Ortiz}, \citenamefont {Yin}, \citenamefont {Su}, \citenamefont {Du},
  \citenamefont {Wang}, \citenamefont {Chen}, \citenamefont {Lu}, \citenamefont
  {Ying}, \citenamefont {Wilson}, \citenamefont {Chen}, \citenamefont {Song},\
  and\ \citenamefont {Yuan}}]{Duan_ScienceChina_2021}%
  \BibitemOpen
  \bibfield  {author} {\bibinfo {author} {\bibfnamefont {W.}~\bibnamefont
  {Duan}}, \bibinfo {author} {\bibfnamefont {Z.}~\bibnamefont {Nie}}, \bibinfo
  {author} {\bibfnamefont {S.}~\bibnamefont {Luo}}, \bibinfo {author}
  {\bibfnamefont {F.}~\bibnamefont {Yu}}, \bibinfo {author} {\bibfnamefont
  {B.~R.}\ \bibnamefont {Ortiz}}, \bibinfo {author} {\bibfnamefont
  {L.}~\bibnamefont {Yin}}, \bibinfo {author} {\bibfnamefont {H.}~\bibnamefont
  {Su}}, \bibinfo {author} {\bibfnamefont {F.}~\bibnamefont {Du}}, \bibinfo
  {author} {\bibfnamefont {A.}~\bibnamefont {Wang}}, \bibinfo {author}
  {\bibfnamefont {Y.}~\bibnamefont {Chen}}, \bibinfo {author} {\bibfnamefont
  {X.}~\bibnamefont {Lu}}, \bibinfo {author} {\bibfnamefont {J.}~\bibnamefont
  {Ying}}, \bibinfo {author} {\bibfnamefont {S.~D.}\ \bibnamefont {Wilson}},
  \bibinfo {author} {\bibfnamefont {X.}~\bibnamefont {Chen}}, \bibinfo {author}
  {\bibfnamefont {Y.}~\bibnamefont {Song}},\ and\ \bibinfo {author}
  {\bibfnamefont {H.}~\bibnamefont {Yuan}},\ }\bibfield  {title} {\bibinfo
  {title} {Nodeless superconductivity in the kagome metal
  {${\mathrm{CsV}}_{3}{\mathrm{Sb}}_{5}$}},\ }\href
  {https://doi.org/10.1007/s11433-021-1747-7} {\bibfield  {journal} {\bibinfo
  {journal} {Science China Physics, Mechanics \& Astronomy}\ }\textbf {\bibinfo
  {volume} {64}},\ \bibinfo {pages} {107462} (\bibinfo {year}
  {2021})}\BibitemShut {NoStop}%
\bibitem [{\citenamefont {Gupta}\ \emph {et~al.}(2022)\citenamefont {Gupta},
  \citenamefont {Das}, \citenamefont {Mielke~III}, \citenamefont {Guguchia},
  \citenamefont {Shiroka}, \citenamefont {Baines}, \citenamefont {Bartkowiak},
  \citenamefont {Luetkens}, \citenamefont {Khasanov}, \citenamefont {Yin},
  \citenamefont {Tu}, \citenamefont {Gong},\ and\ \citenamefont
  {Lei}}]{Gupta_npjQM_2022}%
  \BibitemOpen
  \bibfield  {author} {\bibinfo {author} {\bibfnamefont {R.}~\bibnamefont
  {Gupta}}, \bibinfo {author} {\bibfnamefont {D.}~\bibnamefont {Das}}, \bibinfo
  {author} {\bibfnamefont {C.~H.}\ \bibnamefont {Mielke~III}}, \bibinfo
  {author} {\bibfnamefont {Z.}~\bibnamefont {Guguchia}}, \bibinfo {author}
  {\bibfnamefont {T.}~\bibnamefont {Shiroka}}, \bibinfo {author} {\bibfnamefont
  {C.}~\bibnamefont {Baines}}, \bibinfo {author} {\bibfnamefont
  {M.}~\bibnamefont {Bartkowiak}}, \bibinfo {author} {\bibfnamefont
  {H.}~\bibnamefont {Luetkens}}, \bibinfo {author} {\bibfnamefont
  {R.}~\bibnamefont {Khasanov}}, \bibinfo {author} {\bibfnamefont
  {Q.}~\bibnamefont {Yin}}, \bibinfo {author} {\bibfnamefont {Z.}~\bibnamefont
  {Tu}}, \bibinfo {author} {\bibfnamefont {C.}~\bibnamefont {Gong}},\ and\
  \bibinfo {author} {\bibfnamefont {H.}~\bibnamefont {Lei}},\ }\bibfield
  {title} {\bibinfo {title} {Microscopic evidence for anisotropic multigap
  superconductivity in the {${\mathrm{CsV}}_{3}{\mathrm{Sb}}_{5}$} kagome
  superconductor},\ }\href {https://doi.org/10.1038/s41535-022-00453-7}
  {\bibfield  {journal} {\bibinfo  {journal} {npj Quantum Materials}\ }\textbf
  {\bibinfo {volume} {7}},\ \bibinfo {pages} {49} (\bibinfo {year}
  {2022})}\BibitemShut {NoStop}%
\bibitem [{\citenamefont {Chen}\ \emph
  {et~al.}(2021{\natexlab{a}})\citenamefont {Chen}, \citenamefont {Yang},
  \citenamefont {Hu}, \citenamefont {Zhao}, \citenamefont {Yuan}, \citenamefont
  {Xing}, \citenamefont {Qian}, \citenamefont {Huang}, \citenamefont {Li},
  \citenamefont {Ye}, \citenamefont {Ma}, \citenamefont {Ni}, \citenamefont
  {Zhang}, \citenamefont {Yin}, \citenamefont {Gong}, \citenamefont {Tu},
  \citenamefont {Lei}, \citenamefont {Tan}, \citenamefont {Zhou}, \citenamefont
  {Shen}, \citenamefont {Dong}, \citenamefont {Yan}, \citenamefont {Wang},\
  and\ \citenamefont {Gao}}]{Chen_Nature_2021}%
  \BibitemOpen
  \bibfield  {author} {\bibinfo {author} {\bibfnamefont {H.}~\bibnamefont
  {Chen}}, \bibinfo {author} {\bibfnamefont {H.}~\bibnamefont {Yang}}, \bibinfo
  {author} {\bibfnamefont {B.}~\bibnamefont {Hu}}, \bibinfo {author}
  {\bibfnamefont {Z.}~\bibnamefont {Zhao}}, \bibinfo {author} {\bibfnamefont
  {J.}~\bibnamefont {Yuan}}, \bibinfo {author} {\bibfnamefont {Y.}~\bibnamefont
  {Xing}}, \bibinfo {author} {\bibfnamefont {G.}~\bibnamefont {Qian}}, \bibinfo
  {author} {\bibfnamefont {Z.}~\bibnamefont {Huang}}, \bibinfo {author}
  {\bibfnamefont {G.}~\bibnamefont {Li}}, \bibinfo {author} {\bibfnamefont
  {Y.}~\bibnamefont {Ye}}, \bibinfo {author} {\bibfnamefont {S.}~\bibnamefont
  {Ma}}, \bibinfo {author} {\bibfnamefont {S.}~\bibnamefont {Ni}}, \bibinfo
  {author} {\bibfnamefont {H.}~\bibnamefont {Zhang}}, \bibinfo {author}
  {\bibfnamefont {Q.}~\bibnamefont {Yin}}, \bibinfo {author} {\bibfnamefont
  {C.}~\bibnamefont {Gong}}, \bibinfo {author} {\bibfnamefont {Z.}~\bibnamefont
  {Tu}}, \bibinfo {author} {\bibfnamefont {H.}~\bibnamefont {Lei}}, \bibinfo
  {author} {\bibfnamefont {H.}~\bibnamefont {Tan}}, \bibinfo {author}
  {\bibfnamefont {S.}~\bibnamefont {Zhou}}, \bibinfo {author} {\bibfnamefont
  {C.}~\bibnamefont {Shen}}, \bibinfo {author} {\bibfnamefont {X.}~\bibnamefont
  {Dong}}, \bibinfo {author} {\bibfnamefont {B.}~\bibnamefont {Yan}}, \bibinfo
  {author} {\bibfnamefont {Z.}~\bibnamefont {Wang}},\ and\ \bibinfo {author}
  {\bibfnamefont {H.-J.}\ \bibnamefont {Gao}},\ }\bibfield  {title} {\bibinfo
  {title} {Roton pair density wave in a strong-coupling kagome
  superconductor},\ }\href {https://doi.org/10.1038/s41586-021-03983-5}
  {\bibfield  {journal} {\bibinfo  {journal} {Nature}\ }\textbf {\bibinfo
  {volume} {599}},\ \bibinfo {pages} {222} (\bibinfo {year}
  {2021}{\natexlab{a}})}\BibitemShut {NoStop}%
\bibitem [{\citenamefont {Frachet}\ \emph {et~al.}(2022)\citenamefont
  {Frachet}, \citenamefont {Wiecki}, \citenamefont {Lacmann}, \citenamefont
  {Souliou}, \citenamefont {Willa}, \citenamefont {Meingast}, \citenamefont
  {Merz}, \citenamefont {Haghighirad}, \citenamefont {Le~Tacon},\ and\
  \citenamefont {Böhmer}}]{Frachet_npjQM_2022}%
  \BibitemOpen
  \bibfield  {author} {\bibinfo {author} {\bibfnamefont {M.}~\bibnamefont
  {Frachet}}, \bibinfo {author} {\bibfnamefont {P.}~\bibnamefont {Wiecki}},
  \bibinfo {author} {\bibfnamefont {T.}~\bibnamefont {Lacmann}}, \bibinfo
  {author} {\bibfnamefont {S.~M.}\ \bibnamefont {Souliou}}, \bibinfo {author}
  {\bibfnamefont {K.}~\bibnamefont {Willa}}, \bibinfo {author} {\bibfnamefont
  {C.}~\bibnamefont {Meingast}}, \bibinfo {author} {\bibfnamefont
  {M.}~\bibnamefont {Merz}}, \bibinfo {author} {\bibfnamefont {A.-A.}\
  \bibnamefont {Haghighirad}}, \bibinfo {author} {\bibfnamefont
  {M.}~\bibnamefont {Le~Tacon}},\ and\ \bibinfo {author} {\bibfnamefont
  {A.~E.}\ \bibnamefont {Böhmer}},\ }\bibfield  {title} {\bibinfo {title}
  {Elastoresistivity in the incommensurate charge density wave phase of
  {${\mathrm{BaNi}}_{2}({\mathrm{As}}_{1-x}{\mathrm{P}}_{x})_{2}$}},\ }\href
  {https://doi.org/10.1038/s41535-022-00525-8} {\bibfield  {journal} {\bibinfo
  {journal} {npj Quantum Mater.}\ }\textbf {\bibinfo {volume} {7}},\ \bibinfo
  {pages} {115} (\bibinfo {year} {2022})}\BibitemShut {NoStop}%
\bibitem [{\citenamefont {Kuo}\ \emph {et~al.}(2013)\citenamefont {Kuo},
  \citenamefont {Shapiro}, \citenamefont {Riggs},\ and\ \citenamefont
  {Fisher}}]{Kuo_PRB_2013}%
  \BibitemOpen
  \bibfield  {author} {\bibinfo {author} {\bibfnamefont {H.-H.}\ \bibnamefont
  {Kuo}}, \bibinfo {author} {\bibfnamefont {M.~C.}\ \bibnamefont {Shapiro}},
  \bibinfo {author} {\bibfnamefont {S.~C.}\ \bibnamefont {Riggs}},\ and\
  \bibinfo {author} {\bibfnamefont {I.~R.}\ \bibnamefont {Fisher}},\ }\bibfield
   {title} {\bibinfo {title} {Measurement of the elastoresistivity coefficients
  of the underdoped iron arsenide
  {Ba(Fe$_{0.975}$Co$_{0.025}$)$_{2}$As$_{2}$}},\ }\href
  {https://doi.org/10.1103/PhysRevB.88.085113} {\bibfield  {journal} {\bibinfo
  {journal} {Phys. Rev. B}\ }\textbf {\bibinfo {volume} {88}},\ \bibinfo
  {pages} {085113} (\bibinfo {year} {2013})}\BibitemShut {NoStop}%
\bibitem [{\citenamefont {Zheng}\ \emph {et~al.}(2022)\citenamefont {Zheng},
  \citenamefont {Wu.~Zhimian}, \citenamefont {Nie}, \citenamefont {Shan},
  \citenamefont {Sun}, \citenamefont {Song}, \citenamefont {Yu}, \citenamefont
  {Li}, \citenamefont {Zhao}, \citenamefont {Li}, \citenamefont {Kang},
  \citenamefont {Zhou}, \citenamefont {Liu}, \citenamefont {Xiang},
  \citenamefont {Ying}, \citenamefont {Wang}, \citenamefont {Wu},\ and\
  \citenamefont {Chen}}]{Zheng_Nature_2022}%
  \BibitemOpen
  \bibfield  {author} {\bibinfo {author} {\bibfnamefont {L.}~\bibnamefont
  {Zheng}}, \bibinfo {author} {\bibfnamefont {Y.}~\bibnamefont {Wu.~Zhimian},
  \bibfnamefont {Yang}}, \bibinfo {author} {\bibfnamefont {L.}~\bibnamefont
  {Nie}}, \bibinfo {author} {\bibfnamefont {M.}~\bibnamefont {Shan}}, \bibinfo
  {author} {\bibfnamefont {K.}~\bibnamefont {Sun}}, \bibinfo {author}
  {\bibfnamefont {D.}~\bibnamefont {Song}}, \bibinfo {author} {\bibfnamefont
  {F.}~\bibnamefont {Yu}}, \bibinfo {author} {\bibfnamefont {J.}~\bibnamefont
  {Li}}, \bibinfo {author} {\bibfnamefont {D.}~\bibnamefont {Zhao}}, \bibinfo
  {author} {\bibfnamefont {S.}~\bibnamefont {Li}}, \bibinfo {author}
  {\bibfnamefont {B.}~\bibnamefont {Kang}}, \bibinfo {author} {\bibfnamefont
  {Y.}~\bibnamefont {Zhou}}, \bibinfo {author} {\bibfnamefont {K.}~\bibnamefont
  {Liu}}, \bibinfo {author} {\bibfnamefont {Z.}~\bibnamefont {Xiang}}, \bibinfo
  {author} {\bibfnamefont {J.}~\bibnamefont {Ying}}, \bibinfo {author}
  {\bibfnamefont {Z.}~\bibnamefont {Wang}}, \bibinfo {author} {\bibfnamefont
  {T.}~\bibnamefont {Wu}},\ and\ \bibinfo {author} {\bibfnamefont
  {X.}~\bibnamefont {Chen}},\ }\bibfield  {title} {\bibinfo {title} {Emergent
  charge order in pressurized kagome superconductor
  {$\mathrm{CsV}_{3}\mathrm{Sb}_{5}$}},\ }\href
  {https://doi.org/10.1038/s41586-022-05351-3} {\bibfield  {journal} {\bibinfo
  {journal} {Nature}\ }\textbf {\bibinfo {volume} {611}},\ \bibinfo {pages}
  {682} (\bibinfo {year} {2022})}\BibitemShut {NoStop}%
\bibitem [{\citenamefont {Chen}\ \emph
  {et~al.}(2021{\natexlab{b}})\citenamefont {Chen}, \citenamefont {Wang},
  \citenamefont {Yin}, \citenamefont {Gu}, \citenamefont {Jiang}, \citenamefont
  {Tu}, \citenamefont {Gong}, \citenamefont {Uwatoko}, \citenamefont {Sun},
  \citenamefont {Lei}, \citenamefont {Hu},\ and\ \citenamefont
  {Cheng}}]{Chen_PRL_2021}%
  \BibitemOpen
  \bibfield  {author} {\bibinfo {author} {\bibfnamefont {K.~Y.}\ \bibnamefont
  {Chen}}, \bibinfo {author} {\bibfnamefont {N.~N.}\ \bibnamefont {Wang}},
  \bibinfo {author} {\bibfnamefont {Q.~W.}\ \bibnamefont {Yin}}, \bibinfo
  {author} {\bibfnamefont {Y.~H.}\ \bibnamefont {Gu}}, \bibinfo {author}
  {\bibfnamefont {K.}~\bibnamefont {Jiang}}, \bibinfo {author} {\bibfnamefont
  {Z.~J.}\ \bibnamefont {Tu}}, \bibinfo {author} {\bibfnamefont {C.~S.}\
  \bibnamefont {Gong}}, \bibinfo {author} {\bibfnamefont {Y.}~\bibnamefont
  {Uwatoko}}, \bibinfo {author} {\bibfnamefont {J.~P.}\ \bibnamefont {Sun}},
  \bibinfo {author} {\bibfnamefont {H.~C.}\ \bibnamefont {Lei}}, \bibinfo
  {author} {\bibfnamefont {J.~P.}\ \bibnamefont {Hu}},\ and\ \bibinfo {author}
  {\bibfnamefont {J.-G.}\ \bibnamefont {Cheng}},\ }\bibfield  {title} {\bibinfo
  {title} {Double superconducting dome and triple enhancement of {${T}_{c}$} in
  the kagome superconductor {${\mathrm{CsV}}_{3}{\mathrm{Sb}}_{5}$} under high
  pressure},\ }\href {https://doi.org/10.1103/PhysRevLett.126.247001}
  {\bibfield  {journal} {\bibinfo  {journal} {Phys. Rev. Lett.}\ }\textbf
  {\bibinfo {volume} {126}},\ \bibinfo {pages} {247001} (\bibinfo {year}
  {2021}{\natexlab{b}})}\BibitemShut {NoStop}%
\bibitem [{\citenamefont {Ritz}\ \emph
  {et~al.}(2023{\natexlab{a}})\citenamefont {Ritz}, \citenamefont {Fernandes},\
  and\ \citenamefont {Birol}}]{Turan_PRB_2023}%
  \BibitemOpen
  \bibfield  {author} {\bibinfo {author} {\bibfnamefont {E.~T.}\ \bibnamefont
  {Ritz}}, \bibinfo {author} {\bibfnamefont {R.~M.}\ \bibnamefont
  {Fernandes}},\ and\ \bibinfo {author} {\bibfnamefont {T.}~\bibnamefont
  {Birol}},\ }\bibfield  {title} {\bibinfo {title} {Impact of sb degrees of
  freedom on the charge density wave phase diagram of the kagome metal
  {${\mathrm{CsV}}_{3}{\mathrm{Sb}}_{5}$}},\ }\href
  {https://doi.org/10.1103/PhysRevB.107.205131} {\bibfield  {journal} {\bibinfo
   {journal} {Phys. Rev. B}\ }\textbf {\bibinfo {volume} {107}},\ \bibinfo
  {pages} {205131} (\bibinfo {year} {2023}{\natexlab{a}})}\BibitemShut
  {NoStop}%
\bibitem [{\citenamefont {Ritz}\ \emph
  {et~al.}(2023{\natexlab{b}})\citenamefont {Ritz}, \citenamefont {Røising},
  \citenamefont {Christensen}, \citenamefont {Birol}, \citenamefont
  {Andersen},\ and\ \citenamefont {Fernandes}}]{Ritz_Arxiv_2023}%
  \BibitemOpen
  \bibfield  {author} {\bibinfo {author} {\bibfnamefont {E.}~\bibnamefont
  {Ritz}}, \bibinfo {author} {\bibfnamefont {H.~S.}\ \bibnamefont {Røising}},
  \bibinfo {author} {\bibfnamefont {M.~H.}\ \bibnamefont {Christensen}},
  \bibinfo {author} {\bibfnamefont {T.}~\bibnamefont {Birol}}, \bibinfo
  {author} {\bibfnamefont {B.~M.}\ \bibnamefont {Andersen}},\ and\ \bibinfo
  {author} {\bibfnamefont {R.~M.}\ \bibnamefont {Fernandes}},\ }\href@noop {}
  {\bibinfo {title} {Superconductivity from orbital-selective electron-phonon
  coupling in {$A\mathrm{V}_3\mathrm{Sb}_5$}}} (\bibinfo {year}
  {2023}{\natexlab{b}}),\ \Eprint {https://arxiv.org/abs/2304.14822}
  {arXiv:2304.14822 [cond-mat.supr-con]} \BibitemShut {NoStop}%
\bibitem [{\citenamefont {Qian}\ \emph {et~al.}(2021)\citenamefont {Qian},
  \citenamefont {Christensen}, \citenamefont {Hu}, \citenamefont {Saha},
  \citenamefont {Andersen}, \citenamefont {Fernandes}, \citenamefont {Birol},\
  and\ \citenamefont {Ni}}]{Nie_PRB_2021}%
  \BibitemOpen
  \bibfield  {author} {\bibinfo {author} {\bibfnamefont {T.}~\bibnamefont
  {Qian}}, \bibinfo {author} {\bibfnamefont {M.~H.}\ \bibnamefont
  {Christensen}}, \bibinfo {author} {\bibfnamefont {C.}~\bibnamefont {Hu}},
  \bibinfo {author} {\bibfnamefont {A.}~\bibnamefont {Saha}}, \bibinfo {author}
  {\bibfnamefont {B.~M.}\ \bibnamefont {Andersen}}, \bibinfo {author}
  {\bibfnamefont {R.~M.}\ \bibnamefont {Fernandes}}, \bibinfo {author}
  {\bibfnamefont {T.}~\bibnamefont {Birol}},\ and\ \bibinfo {author}
  {\bibfnamefont {N.}~\bibnamefont {Ni}},\ }\bibfield  {title} {\bibinfo
  {title} {Revealing the competition between charge density wave and
  superconductivity in {${{\mathrm{CsV}}_{3}\mathrm{Sb}}_{5}$} through uniaxial
  strain},\ }\href {https://doi.org/10.1103/PhysRevB.104.144506} {\bibfield
  {journal} {\bibinfo  {journal} {Phys. Rev. B}\ }\textbf {\bibinfo {volume}
  {104}},\ \bibinfo {pages} {144506} (\bibinfo {year} {2021})}\BibitemShut
  {NoStop}%
\bibitem [{\citenamefont {Consiglio}\ \emph {et~al.}(2022)\citenamefont
  {Consiglio}, \citenamefont {Schwemmer}, \citenamefont {Wu}, \citenamefont
  {Hanke}, \citenamefont {Neupert}, \citenamefont {Thomale}, \citenamefont
  {Sangiovanni},\ and\ \citenamefont {Di~Sante}}]{Consiglio_PRB_2022}%
  \BibitemOpen
  \bibfield  {author} {\bibinfo {author} {\bibfnamefont {A.}~\bibnamefont
  {Consiglio}}, \bibinfo {author} {\bibfnamefont {T.}~\bibnamefont
  {Schwemmer}}, \bibinfo {author} {\bibfnamefont {X.}~\bibnamefont {Wu}},
  \bibinfo {author} {\bibfnamefont {W.}~\bibnamefont {Hanke}}, \bibinfo
  {author} {\bibfnamefont {T.}~\bibnamefont {Neupert}}, \bibinfo {author}
  {\bibfnamefont {R.}~\bibnamefont {Thomale}}, \bibinfo {author} {\bibfnamefont
  {G.}~\bibnamefont {Sangiovanni}},\ and\ \bibinfo {author} {\bibfnamefont
  {D.}~\bibnamefont {Di~Sante}},\ }\bibfield  {title} {\bibinfo {title} {Van
  hove tuning of {A${\mathrm{V}}_{3}{\mathrm{Sb}}_{5}$} kagome metals under
  pressure and strain},\ }\href {https://doi.org/10.1103/PhysRevB.105.165146}
  {\bibfield  {journal} {\bibinfo  {journal} {Phys. Rev. B}\ }\textbf {\bibinfo
  {volume} {105}},\ \bibinfo {pages} {165146} (\bibinfo {year}
  {2022})}\BibitemShut {NoStop}%
\bibitem [{\citenamefont {Böhmer}\ \emph {et~al.}(2015)\citenamefont
  {Böhmer}, \citenamefont {Hardy}, \citenamefont {Wang}, \citenamefont {Wolf},
  \citenamefont {Schweiss},\ and\ \citenamefont
  {Meingast}}]{Bohmer_NatComm_2015}%
  \BibitemOpen
  \bibfield  {author} {\bibinfo {author} {\bibfnamefont {A.~E.}\ \bibnamefont
  {Böhmer}}, \bibinfo {author} {\bibfnamefont {F.}~\bibnamefont {Hardy}},
  \bibinfo {author} {\bibfnamefont {L.}~\bibnamefont {Wang}}, \bibinfo {author}
  {\bibfnamefont {T.}~\bibnamefont {Wolf}}, \bibinfo {author} {\bibfnamefont
  {P.}~\bibnamefont {Schweiss}},\ and\ \bibinfo {author} {\bibfnamefont
  {C.}~\bibnamefont {Meingast}},\ }\bibfield  {title} {\bibinfo {title}
  {Superconductivity-induced re-entrance of the orthorhombic distortion in
  {$\mathrm{Ba}_{1-x}\mathrm{K}_x\mathrm{Fe}_{2}\mathrm{As}_{2}$}},\ }\href
  {https://doi.org/10.1038/ncomms8911} {\bibfield  {journal} {\bibinfo
  {journal} {Nature Communications}\ }\textbf {\bibinfo {volume} {6}},\
  \bibinfo {pages} {7911} (\bibinfo {year} {2015})}\BibitemShut {NoStop}%
\bibitem [{\citenamefont {Wang}\ \emph {et~al.}(2016)\citenamefont {Wang},
  \citenamefont {Hardy}, \citenamefont {B\"ohmer}, \citenamefont {Wolf},
  \citenamefont {Schweiss},\ and\ \citenamefont {Meingast}}]{Wang_PRB_93}%
  \BibitemOpen
  \bibfield  {author} {\bibinfo {author} {\bibfnamefont {L.}~\bibnamefont
  {Wang}}, \bibinfo {author} {\bibfnamefont {F.}~\bibnamefont {Hardy}},
  \bibinfo {author} {\bibfnamefont {A.~E.}\ \bibnamefont {B\"ohmer}}, \bibinfo
  {author} {\bibfnamefont {T.}~\bibnamefont {Wolf}}, \bibinfo {author}
  {\bibfnamefont {P.}~\bibnamefont {Schweiss}},\ and\ \bibinfo {author}
  {\bibfnamefont {C.}~\bibnamefont {Meingast}},\ }\bibfield  {title} {\bibinfo
  {title} {Complex phase diagram of
  {${\mathrm{Ba}}_{1\ensuremath{-}x}{\mathrm{Na}}_{x}{\mathrm{Fe}}_{2}{\mathrm{As}}_{2}$}:
  A multitude of phases striving for the electronic entropy},\ }\href
  {https://doi.org/10.1103/PhysRevB.93.014514} {\bibfield  {journal} {\bibinfo
  {journal} {Phys. Rev. B}\ }\textbf {\bibinfo {volume} {93}},\ \bibinfo
  {pages} {014514} (\bibinfo {year} {2016})}\BibitemShut {NoStop}%
\bibitem [{\citenamefont {Chu}\ \emph {et~al.}(2012)\citenamefont {Chu},
  \citenamefont {Kuo}, \citenamefont {Analytis},\ and\ \citenamefont
  {Fisher}}]{Chu_Science_2012}%
  \BibitemOpen
  \bibfield  {author} {\bibinfo {author} {\bibfnamefont {J.-H.}\ \bibnamefont
  {Chu}}, \bibinfo {author} {\bibfnamefont {H.-H.}\ \bibnamefont {Kuo}},
  \bibinfo {author} {\bibfnamefont {J.~G.}\ \bibnamefont {Analytis}},\ and\
  \bibinfo {author} {\bibfnamefont {I.~R.}\ \bibnamefont {Fisher}},\ }\bibfield
   {title} {\bibinfo {title} {{Divergent Nematic Susceptibility in an Iron
  Arsenide Superconductor}},\ }\href {https://doi.org/10.1126/science.1221713}
  {\bibfield  {journal} {\bibinfo  {journal} {Science}\ }\textbf {\bibinfo
  {volume} {337}},\ \bibinfo {pages} {710} (\bibinfo {year}
  {2012})}\BibitemShut {NoStop}%
\bibitem [{\citenamefont {Wiecki}\ \emph {et~al.}(2021)\citenamefont {Wiecki},
  \citenamefont {Frachet}, \citenamefont {Haghighirad}, \citenamefont {Wolf},
  \citenamefont {Meingast}, \citenamefont {Heid},\ and\ \citenamefont
  {Böhmer}}]{Wiecki_NatComm_2021}%
  \BibitemOpen
  \bibfield  {author} {\bibinfo {author} {\bibfnamefont {P.}~\bibnamefont
  {Wiecki}}, \bibinfo {author} {\bibfnamefont {M.}~\bibnamefont {Frachet}},
  \bibinfo {author} {\bibfnamefont {A.-A.}\ \bibnamefont {Haghighirad}},
  \bibinfo {author} {\bibfnamefont {T.}~\bibnamefont {Wolf}}, \bibinfo {author}
  {\bibfnamefont {C.}~\bibnamefont {Meingast}}, \bibinfo {author}
  {\bibfnamefont {R.}~\bibnamefont {Heid}},\ and\ \bibinfo {author}
  {\bibfnamefont {A.~E.}\ \bibnamefont {Böhmer}},\ }\bibfield  {title}
  {\bibinfo {title} {Emerging symmetric strain response and weakening nematic
  fluctuations in strongly hole-doped iron-based superconductors},\ }\href
  {https://doi.org/10.1038/s41467-021-25121-5} {\bibfield  {journal} {\bibinfo
  {journal} {Nature Communications}\ }\textbf {\bibinfo {volume} {12}},\
  \bibinfo {pages} {4824} (\bibinfo {year} {2021})}\BibitemShut {NoStop}%
\bibitem [{Note1()}]{Note1}%
  \BibitemOpen
  \bibinfo {note} {Note that our own measurements do also show a downturn below
  $\approx $ 35K when uncorrected for the residual resistivity. See empty
  symbols in Fig.2(a)}\BibitemShut {NoStop}%
\bibitem [{\citenamefont {Liu}\ \emph {et~al.}(2023)\citenamefont {Liu},
  \citenamefont {Shi}, \citenamefont {Jiang}, \citenamefont {Rosenberg},
  \citenamefont {DeStefano}, \citenamefont {Liu}, \citenamefont {Hu},
  \citenamefont {Zhao}, \citenamefont {Wang}, \citenamefont {Yao},
  \citenamefont {Graf}, \citenamefont {Dai}, \citenamefont {Yang},
  \citenamefont {Xu},\ and\ \citenamefont {Chu}}]{liu2023}%
  \BibitemOpen
  \bibfield  {author} {\bibinfo {author} {\bibfnamefont {Z.}~\bibnamefont
  {Liu}}, \bibinfo {author} {\bibfnamefont {Y.}~\bibnamefont {Shi}}, \bibinfo
  {author} {\bibfnamefont {Q.}~\bibnamefont {Jiang}}, \bibinfo {author}
  {\bibfnamefont {E.~W.}\ \bibnamefont {Rosenberg}}, \bibinfo {author}
  {\bibfnamefont {J.~M.}\ \bibnamefont {DeStefano}}, \bibinfo {author}
  {\bibfnamefont {J.}~\bibnamefont {Liu}}, \bibinfo {author} {\bibfnamefont
  {C.}~\bibnamefont {Hu}}, \bibinfo {author} {\bibfnamefont {Y.}~\bibnamefont
  {Zhao}}, \bibinfo {author} {\bibfnamefont {Z.}~\bibnamefont {Wang}}, \bibinfo
  {author} {\bibfnamefont {Y.}~\bibnamefont {Yao}}, \bibinfo {author}
  {\bibfnamefont {D.}~\bibnamefont {Graf}}, \bibinfo {author} {\bibfnamefont
  {P.}~\bibnamefont {Dai}}, \bibinfo {author} {\bibfnamefont {J.}~\bibnamefont
  {Yang}}, \bibinfo {author} {\bibfnamefont {X.}~\bibnamefont {Xu}},\ and\
  \bibinfo {author} {\bibfnamefont {J.-H.}\ \bibnamefont {Chu}},\ }\href@noop
  {} {\bibinfo {title} {Absence of nematic instability in the kagome metal
  csv$_3$sb$_5$}} (\bibinfo {year} {2023}),\ \Eprint
  {https://arxiv.org/abs/2309.14574} {arXiv:2309.14574 [cond-mat.supr-con]}
  \BibitemShut {NoStop}%
\bibitem [{\citenamefont {Asaba}\ \emph {et~al.}(2023)\citenamefont {Asaba},
  \citenamefont {Onishi}, \citenamefont {Kageyama}, \citenamefont {Kiyosue},
  \citenamefont {Ohtsuka}, \citenamefont {Suetsugu}, \citenamefont {Kohsaka},
  \citenamefont {Gaggl}, \citenamefont {Kasahara}, \citenamefont {Murayama},
  \citenamefont {Hashimoto}, \citenamefont {Tazai}, \citenamefont {Kontani},
  \citenamefont {Ortiz}, \citenamefont {Wilson}, \citenamefont {Li},
  \citenamefont {Wen}, \citenamefont {Shibauchi},\ and\ \citenamefont
  {Matsuda}}]{asaba2023}%
  \BibitemOpen
  \bibfield  {author} {\bibinfo {author} {\bibfnamefont {T.}~\bibnamefont
  {Asaba}}, \bibinfo {author} {\bibfnamefont {A.}~\bibnamefont {Onishi}},
  \bibinfo {author} {\bibfnamefont {Y.}~\bibnamefont {Kageyama}}, \bibinfo
  {author} {\bibfnamefont {T.}~\bibnamefont {Kiyosue}}, \bibinfo {author}
  {\bibfnamefont {K.}~\bibnamefont {Ohtsuka}}, \bibinfo {author} {\bibfnamefont
  {S.}~\bibnamefont {Suetsugu}}, \bibinfo {author} {\bibfnamefont
  {Y.}~\bibnamefont {Kohsaka}}, \bibinfo {author} {\bibfnamefont
  {T.}~\bibnamefont {Gaggl}}, \bibinfo {author} {\bibfnamefont
  {Y.}~\bibnamefont {Kasahara}}, \bibinfo {author} {\bibfnamefont
  {H.}~\bibnamefont {Murayama}}, \bibinfo {author} {\bibfnamefont
  {K.}~\bibnamefont {Hashimoto}}, \bibinfo {author} {\bibfnamefont
  {R.}~\bibnamefont {Tazai}}, \bibinfo {author} {\bibfnamefont
  {H.}~\bibnamefont {Kontani}}, \bibinfo {author} {\bibfnamefont {B.~R.}\
  \bibnamefont {Ortiz}}, \bibinfo {author} {\bibfnamefont {S.~D.}\ \bibnamefont
  {Wilson}}, \bibinfo {author} {\bibfnamefont {Q.}~\bibnamefont {Li}}, \bibinfo
  {author} {\bibfnamefont {H.~H.}\ \bibnamefont {Wen}}, \bibinfo {author}
  {\bibfnamefont {T.}~\bibnamefont {Shibauchi}},\ and\ \bibinfo {author}
  {\bibfnamefont {Y.}~\bibnamefont {Matsuda}},\ }\href@noop {} {\bibinfo
  {title} {Evidence for an odd-parity nematic phase above the charge density
  wave transition in kagome metal csv$_3$sb$_5$}} (\bibinfo {year} {2023}),\
  \Eprint {https://arxiv.org/abs/2309.16985} {arXiv:2309.16985
  [cond-mat.str-el]} \BibitemShut {NoStop}%
\end{thebibliography}%

\end{document}


\title{Supplementary Materials: Colossal \textit{c}-axis response and lack of rotational symmetry breaking within the kagome plane of the \Cs~superconductor}

\author{Mehdi Frachet}
 \email{mehdi.frachet@gmail.com}
 \affiliation{Institute for Quantum Materials and Technologies, Karlsruhe Institute of Technology, D-76021 Karlsruhe, Germany}

\author{Liran Wang}
 \affiliation{Institute for Quantum Materials and Technologies, Karlsruhe Institute of Technology, D-76021 Karlsruhe, Germany}

\author{Wei Xia}
 \affiliation{School of Physical Science and Technology, ShanghaiTech University, Shanghai 201210, China}
 \affiliation{ShanghaiTech Laboratory for Topological Physics, Shanghai 201210, China}

\author{Yanfeng Guo}
 \affiliation{School of Physical Science and Technology, ShanghaiTech University, Shanghai 201210, China}
 \affiliation{ShanghaiTech Laboratory for Topological Physics, Shanghai 201210, China}
 
 \author{Mingquan He}
 \affiliation{Low Temperature Physics Laboratory, College of Physics \& Center of Quantum Materials and Devices,
Chongqing University, Chongqing 401331, China}

\author{Nour Maraytta}
 \affiliation{Institute for Quantum Materials and Technologies, Karlsruhe Institute of Technology, D-76021 Karlsruhe, Germany}

\author{Rolf Heid}
 \affiliation{Institute for Quantum Materials and Technologies, Karlsruhe Institute of Technology, D-76021 Karlsruhe, Germany}

\author{Amir-Abbas Haghighirad}
 \affiliation{Institute for Quantum Materials and Technologies, Karlsruhe Institute of Technology, D-76021 Karlsruhe, Germany}

\author{Michael Merz}
 \affiliation{Institute for Quantum Materials and Technologies, Karlsruhe Institute of Technology, D-76021 Karlsruhe, Germany}
 \affiliation{Karlsruhe Nano Micro Facility (KNMFi), Karlsruhe Institute of Technology, 76344 Eggenstein-Leopoldshafen\\}

\author{Christoph Meingast}
 \email{christoph.meingast@kit.edu}
 \affiliation{Institute for Quantum Materials and Technologies, Karlsruhe Institute of Technology, D-76021 Karlsruhe, Germany}

\author{Frédéric Hardy}
 \email{frederic.hardy@kit.edu}
 \affiliation{Institute for Quantum Materials and Technologies, Karlsruhe Institute of Technology, D-76021 Karlsruhe, Germany}

\date{\today}

\begin{abstract}
\end{abstract}

\maketitle

\section{Samples characterization}

\subsection{Freestanding electrical resistance}
\begin{figure*}[h]
\centering
\includegraphics[width=17.2cm]{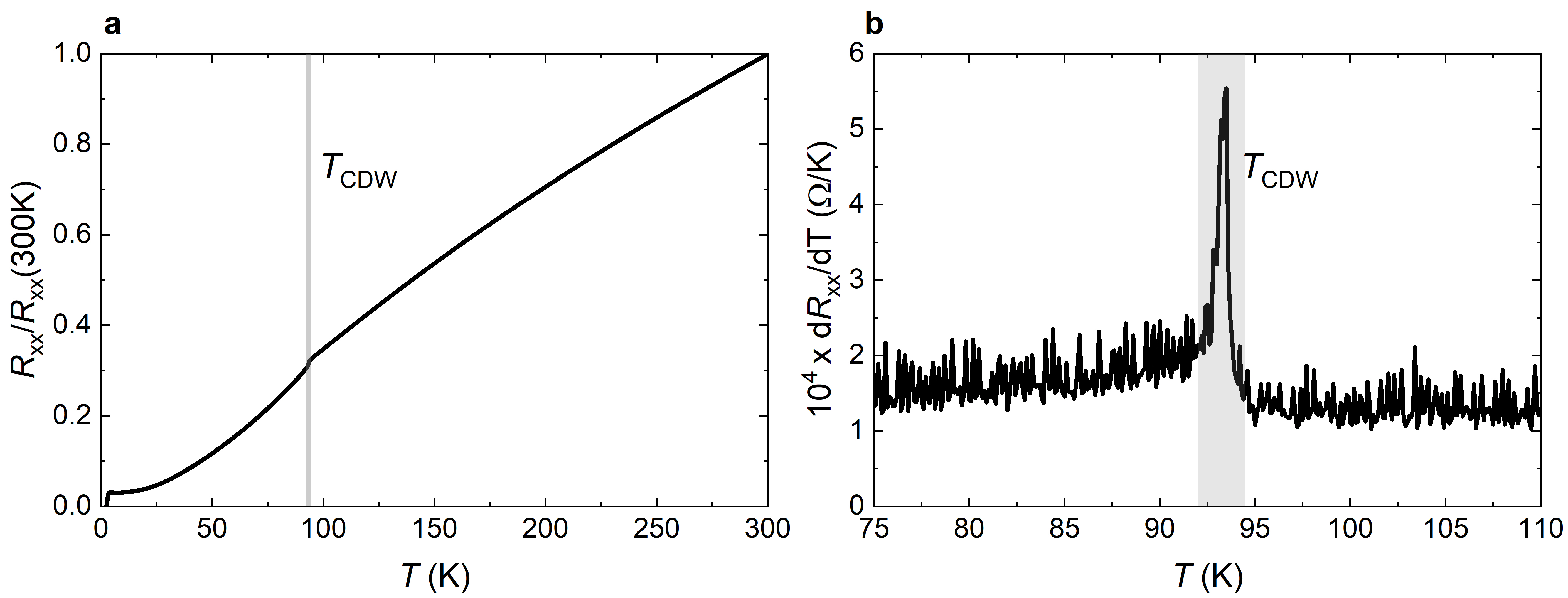}
\caption{(a) Room-temperature normalized in-plane resistance of \Cs~(on warming). In addition to the superconducting transition below $\approx$ 3~K, an anomaly is resolved at the CDW transition at $T_{\rm{CDW}} \approx 94~$K. \textbf{(b)} Zoom around the CDW transition, where $dR_{xx}/dT$ shows a sharp maximum.}
\label{freestanding_resistance}
\end{figure*}

We report the freestanding electrical resistance of a sample from batch B (the batch used for our elastoresistance measurement and $c-$axis thermal expansion) in Fig. \ref{freestanding_resistance}. The residual resistivity ratio (RRR) amounts to $\approx 30$, and a sharp resistance drop occurs on cooling across $T_{\rm{CDW}} \approx 94~$K, that signals the first-order CDW transition. The thermal hysteresis associated to this transition is minute, and is at most $\approx 0.5~$K (not shown).\\

\subsection{Energy dispersive x-ray spectroscopy }
\begin{figure*}[h]
\centering
\includegraphics[width=17.2cm]{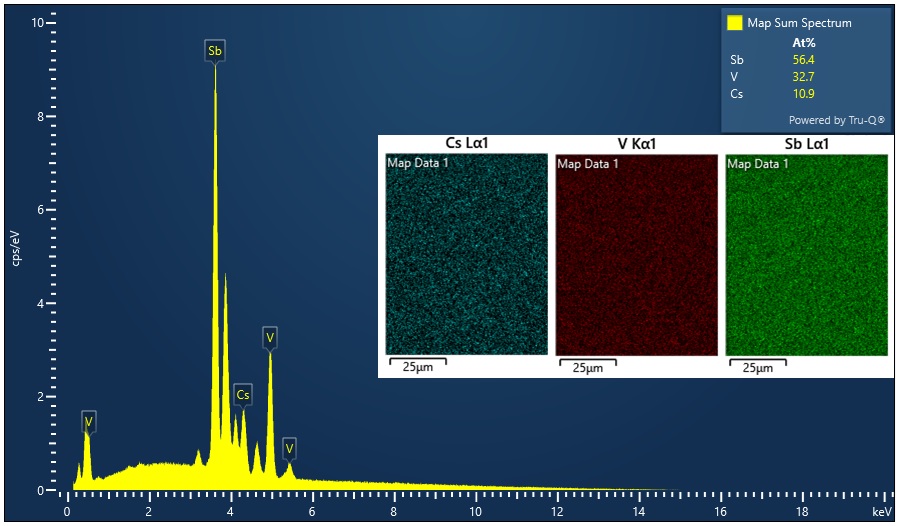}
\caption{EDS spectrum for samples from Batch B. The inset shows the elemental mapping.}
\label{EDX}
\end{figure*}

Chemical composition of CsV$_3$Sb$_5$ crystals was performed using energy dispersive x-ray spectroscopy (EDS) in a COXEM EM-30plus electron microscope equipped with an Oxford Silicon-Drift-Detector (SDD) and AZtecLiveLite-software package. The EDS revealed a quantitative elemental composition of Cs:V:Sb 1:3:5.17. The EDS spectrum and the elemental mapping (inset) are shown in Fig.\ref{EDX}.

\vspace{10mm}

\subsection{XRD}

X-ray diffraction (XRD) data on representative CsV$_3$Sb$_5$ single crystals from the same batches were collected at 295 K on a STOE imaging plate diffraction system (IPDS-2T) using Mo $K_{\alpha}$ radiation. All accessible reflections ($\approx 4.600$) were measured up to a maximum angle of $2 \Theta =65^{\circ}$.\@ The data were corrected for Lorentz, polarization, extinction, and absorption effects. Using SHELXL~\cite{Sheldrick:fa3356} and JANA2006~\cite{PetricekDusekPalatinus+2014+345+352}, all averaged symmetry-independent reflections ($I > 2 \sigma$) have been included for the refinement. The unit cell and the space group (SG) were determined, the atoms were localized in the unit cell utilizing random phases as well as Patterson superposition methods, the structure was completed and solved using difference Fourier analysis, and finally the structure was refined. In all cases the refinements converged quite well and show excellent reliability factors (see GOF, $R_1$,\@ and $wR_2$ in Table \ref{tab:table1}).\@ All of the crystallographic data are in good agreement with the already published ones of Ref.~\cite{Ortiz_PRM_2019}. According to our refinement all Wyckoff positions are completely occupied.

\begin{table}[h]
\caption{\label{tab:table1} Crystallographic data for CsV$_3$Sb$_5$ at 295 K determined from single-crystal x-ray diffraction. The structure was refined in the hexagonal space group (SG) $P6/mmm$.\@ $U_{\rm equiv}$ denotes the equivalent atomic displacement parameters (ADP). The ADPs were refined anisotropically but due to space limitations only the $U_{\rm equiv}$ are listed in the table. The corresponding Wyckoff positions (Wyck.) are given as well.}

\begin{ruledtabular}
		\begin{tabular}[t]{llc}
			& SG                        &     $P6/mmm$          \\
			&  $a$ (\AA)              &      5.5060(7)             \\
			&  $c$ (\AA)              &      9.3274(29)      \\
			&  $\alpha$ (\textdegree)             &       90              \\
			&  $\beta$ (\textdegree)              &       90               \\
			&  $\gamma$ (\textdegree)             &      120             \\
			&  V (\AA$^3$)            &      244.9            \\ 
    Cs      & Wyck.                     &      $1b$                  \\
			& $x$                       &        0                \\
			& $y$                       &        0                    \\
			& $z$                       &        $\frac{1}{2}$                \\
			& $U_{\rm equiv}$ (\AA$^2$) &      0.02991(44)       \\ 
    V       & Wyck.                     &       $3f$              \\
            & $x$                       &        $\frac{1}{2}$         \\
			& $y$                       &          0          \\
			& $z$                       &        0          \\
			& $U_{\rm equiv}$ (\AA$^2$) &  0.01356(61)          \\
    Sb1      & Wyck.                     &      $1a$            \\
            & $x$                       &        0              \\
			& $y$                       &        0                \\
			& $z$                       &        0        \\
			& $U_{\rm equiv}$ (\AA$^2$)   &    0.01532(34)          \\
   Sb2      & Wyck.                     &      $4h$            \\
            & $x$                       &        $\frac{1}{3}$              \\
			& $y$                       &        $\frac{2}{3}$               \\
			& $z$                       &        0.242096        \\
			& $U_{\rm equiv}$ (\AA$^2$)   &    0.01655(14)          \\ \hline
			& GOF                       &       1.23                 \\
			& $wR_2$ (\%)               &       3.93               \\
			& $R_1$ (\%)                &       1.84             \\
		\end{tabular}
\end{ruledtabular}
\end{table}

\pagebreak{}

\section{Theoretical estimate of the elastic constants from DFT}

To obtain rough estimates for the elastic constants we performed ab initio calculations of electronic and vibrational properties of CsV$_3$Sb$_5$ in the framework of the generalized gradient approximation \cite{perde96} using a mixed-basis pseudopotential method \cite{meyer97}.
Phonon dispersions and corresponding interatomic force constants were calculated via density functional perturbation theory (DFTP) \cite{heid99}.
The elastic constants were then derived via the method of long waves, which relates the elastic constants to sums over force constants weighted by the bond vector \cite{born54}.

We employed normconserving pseudopotentials of Vanderbilt type \cite{vande85} and included Cs 5p as well as V 3s and 3p semicore states in the valence space. The mixed-basis set consisted of plane-waves with a cutoff of 24 Ry, augmented by local functions at all atomic sites.
Brillouin zone (BZ) summations were done on a hexagonal
$18\times 18\times 12$ $k$-point mesh (222 points in the irreducible part of the BZ) combined with a Gaussian broadening of 0.1 eV.
Calculations were performed for high-temperature hexagonal structure (space group $P6/mmm$) using experimental room-temperature lattice parameters $a=5.4949$~\AA, $c=9.3085$~\AA) \cite{Ortiz_PRM_2019}.
The single internal structural parameter, the $z$ parameter of
Sb2, was relaxed until atomic forces were below 2.6~meV/\AA,
resulting in $z=0.74156$ in good agreement with the experimental value ($z=0.74217$ \cite{Ortiz_PRM_2019}).

To extract elastic constants, force constants were calculated via DFPT on a $6\times 6\times 2$ hexagonal momentum mesh.
This gave the following estimates (in units of GPa): $c_{11}=108$, $c_{33}=32$,  $c_{12}=52$,  $c_{13}=16$,  $c_{44}=22$,  $c_{66}=28$.
The corresponding phonon dispersion exhibits the well-known unstable branch related to the charge-density wave transition. Because this instability occurs close to the BZ boundary (near M and L points), it is separated from the linear dispersion of the acoustic modes in the vicinity of the BZ center, which determines the elastic constants. Yet, the derived values for the elastic constants should be considered as crude estimates only.

\pagebreak{}

\section{Poisson ratio and correspondence between single piezoelectric and uniaxial stress cell experiments}
In the hexagonal structure of \Cs~the elastic compliance tensor, $\Bar{S}$, is given by \citep{Ballato_1995, Luthi_book},
    
    \[\bar{S} = \begin{bmatrix}
        s_{11} & s_{12} & s_{13} & 0 & 0 & 0 \\
        s_{12} & s_{11} & s_{13} & 0 & 0 & 0 \\
        s_{13} & s_{13} & s_{33} & 0 & 0 & 0 \\
        0 & 0 & 0& s_{44} & 0 & 0 \\
        0 & 0 & 0& 0 & s_{44} & 0 \\
        0 & 0 & 0& 0 & 0 & 2\left(s_{11}-s_{12}\right)
    \end{bmatrix}\] 

And the inverse tensor, the tensor of the elastic constants is given by, 

    \[\bar{C} = \begin{bmatrix}
        c_{11} & c_{12} & c_{13} & 0 & 0 & 0 \\
        c_{12} & c_{11} & c_{13} & 0 & 0 & 0 \\
        c_{13} & c_{13} & c_{33} & 0 & 0 & 0 \\
        0 & 0 & 0& c_{44} & 0 & 0 \\
        0 & 0 & 0& 0 & c_{44} & 0 \\
        0 & 0 & 0& 0 & 0 & \left(c_{11}-c_{12}\right)/2
    \end{bmatrix}\] 

    With indexes given in Voigt notations, where, 

    \begin{equation}
        1 = \rm{xx}, 2 = \rm{yy}, 3=\rm{zz}, 4=\rm{xz}, 6=\rm{xy} 
        \label{Voigt_notation}
    \end{equation}

In particular, 
\begin{equation}
    s_{11} = \frac{1}{2} \left( \frac{c_{33}}{C} + \frac{1}{c_{11}-c_{12}} \right)
\end{equation}

\begin{equation}
    s_{12} = \frac{1}{2} \left( \frac{c_{33}}{C} - \frac{1}{c_{11}-c_{12}} \right)
\end{equation}

\begin{equation}
    s_{13} = \frac{-c_{13}}{C}
\end{equation}

Providing that the constant $C$ is given by,

\begin{equation}
    C = c_{33} (c_{11}+c_{12})-2c_{13}^2
\end{equation}

In turns, the in-plane Poisson ratio relying tension along the $a-$axis to in-plane tranverse compression equals to,

\begin{equation}
    \nu_{\rm{in}}^{\rm{sample}} = s_{12}/s_{11} \approx - 0.44
\end{equation}

Where elastic constants values from DFT are used for the numerical estimation (see \textcolor{red}{ZZ}). Similarly, the out-of-plane sample Poisson ratio equals to,

\begin{equation}
    \nu_{\rm{out}}^{\rm{sample}} = s_{13}/s_{11} \approx - 0.28
\end{equation}

Notably, our  estimation of the in-plane Poisson ratio leads to a value similar to that of the in-plane Poisson ratio of the piezoelectric stack \citep{Kuo_PRB_2013}. This means that, in our experiment, applying tension (or compression) using the piezoelectric stack leads to an essentially similar sample strain state as reached by Qian et al. \citep{Nie_PRB_2021} under uniaxial pressure, up to the ratio of the respective strain transmission. In particular, it means that, 

\begin{equation}
    \frac{d\Tcdw}{dP_{a}} \approx - \frac{d\Tcdw}{d\epsilon_{xx}}
\end{equation}

The left hand-side corresponds to the shift of \Tcdw~under uniaxial pressure along the $a-$axis, where the sample is free to distorts along transverse directions, i.e. the quantity extracted via direct uniaxial stress experiment \citep{Nie_PRB_2021} or from the discontinuity in the thermal expansion at \Tcdw~(table. I in the main text). The right hand-side corresponds to the change of \Tcdw~while the sample is glued to the piezoelectric stack and the transverse in-plane distortion is given by $\nu_{\rm{in}}^{\rm{piezo}}$. The opposite sign comes from the fact that positive pressure corresponds to compression, while positive strain refers to tension. In light of this qualitative correspondence, it is interesting to note that the peak of elastoresistance at \Tcdw~is resolved ref. \citep{Nie_PRB_2021} (see Fig. \textcolor{red}{2} of the mentioned reference), as is the enhanced elastoresistance within the CDW state. Thus, our data qualitatively agree with that of Qian et al.


\section{Definition of the symmetry-resolved elastoresistance coefficients}

On general ground, the elastoresistance coefficients are defined by,

\begin{equation}
    \left(\Delta R/ R\right)_{i} = \sum_{j} m_{ij} \epsilon_{j}
\end{equation}

Where we use again the Voigt notations (see eq. \ref{Voigt_notation}). The tensor of the elastoresistance coefficients is given by,

    \[\bar{m} = \begin{bmatrix}
        m_{11} & m_{12} & m_{13} & 0 & 0 & 0 \\
        m_{12} & m_{11} & m_{13} & 0 & 0 & 0 \\
        m_{13} & m_{13} & m_{33} & 0 & 0 & 0 \\
        0 & 0 & 0& m_{44} & 0 & 0 \\
        0 & 0 & 0& 0 & m_{44} & 0 \\
        0 & 0 & 0& 0 & 0 & \left(m_{11}-m_{12}\right)/2

    \end{bmatrix}\]

Hence, for longitudinal elastoresistance measurements, the resistance variation with the longitudinal strain is,

\begin{equation}
    \left(\Delta R/R\right)_{xx} = m_{11} \epsilon_{xx} + m_{12} \epsilon_{yy} + m_{13} \epsilon_{zz}
\end{equation}

And similarly for the transverse measurements,

\begin{equation}
    \left(\Delta R/R\right)_{yy} = m_{12} \epsilon_{xx} + m_{11} \epsilon_{yy} + m_{13} \epsilon_{zz}
\end{equation}

Using these two equations, and the definition of the Poisson ratios, we have,

\begin{equation}
\begin{aligned}
    \left(\Delta R/R\right)_{xx} - \left(\Delta R/R\right)_{yy} 
    &= \left(m_{11}- m_{12}\right) (\epsilon_{xx}-\epsilon_{yy})\\
    &= \left(m_{11}- m_{12}\right) \left(1-\nu_{\rm{in}}^{\rm{piezo}} \right) \epsilon_{xx}
\end{aligned}
\end{equation}

Similarly, we extract the sum of the resistance variations in transverse and longitudinal geometries, that reads as,

\begin{equation}
\begin{aligned}
    \left(\Delta R/R\right)_{xx} + \left(\Delta R/R\right)_{yy} 
    & = \left( m_{11}+ m_{12} +2m_{13} \frac{\nu_{\rm{out}}^{\rm{sample}}}{1+\nu_{\rm{in}}^{\rm{piezo}}} \right)  (\epsilon_{xx}+\epsilon_{yy})\\
    & = \left( \left(m_{11}+ m_{12}\right) \left(1+\nu_{\rm{in}}^{\rm{piezo}} \right) + 2m_{13} \nu_{\rm{out}}^{\rm{sample}} \right) \epsilon_{xx}
    \end{aligned}
\end{equation}

Thus, in the main text, we simply define,
\begin{equation}
    m_{E_{2g}} \equiv m_{11}- m_{12}
\end{equation}
and, 
\begin{equation}
    m_{A_{1g}} \equiv \left( m_{11}+ m_{12} +2m_{13} \frac{\nu_{\rm{out}}^{\rm{sample}}}{1+\nu_{\rm{in}}^{\rm{piezo}}} \right) 
\end{equation}

That transform as $E_{2g}$ and $A_{1g}$  irreducible representations in the $D_{6h}$ point group, respectively.
Note that  $m_{A_{1g}}$ includes contributions from both in-plane and out-of-plane effects. 


\section{Experimental details on the elastoresistance}
Here, we provide extensive details on the elastoresistance measurements reported in Fig. 2 of the main manuscript. 

\subsection{Zero voltage measurements}

\begin{figure}[h]
\centering
\includegraphics[width=8.6cm]{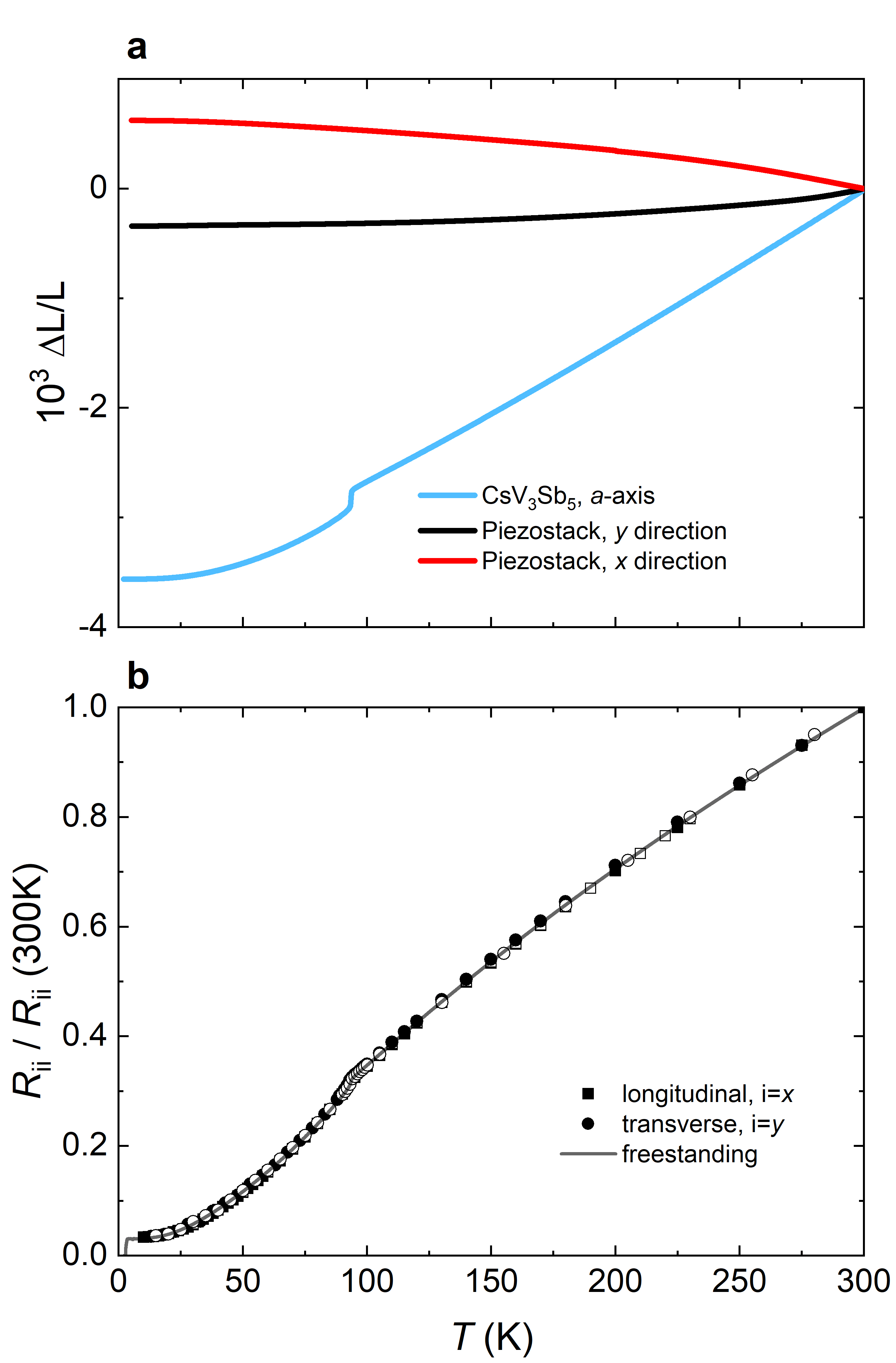}
\caption{\textbf{(a)} Length change, $\Delta L/L$, of the piezoelectric stack along (red) and perpendicular (black) to the poling direction, reproduced from Ref.\citep{Frachet_npjQM_2022}. The length change along the $a-$axis of CsV$_3$Sb$_5$ is taken from Fig. 1 of the main text. The differential thermal expansion between the sample and the piezoelectric induces isotropic and anisotropic strain components, the former being larger. \textbf{(b)} Resistance of \Cs~glued to the piezoelectric stack as a function of temperature in the longitudinal (circles) and transverse (squares) directions. The freestanding resistance is shown for comparison. All curves are normalized by the resistance at 300~K. }
\label{FigResistanceGlued}
\end{figure}

We first describe measurements performed without applying voltages to the piezoelectric stack. Even in that case, strain is applied to the sample because of the thermal expansion mismatch between the sample and the piezoelectric. Since the thermal expansion of the piezoelectric is anisotropic, see Fig. \ref{FigResistanceGlued}a, a (small) anisotropic component of the strain is present \citep{Frachet_npjQM_2022}. In the advent of a 6-fold rotational symmetry breaking, the electrical resistance along and perpendicular to the piezoelectric stack poling axis should in turn develop an anisotropy as the strain partially detwins the different domains \citep{Frachet_npjQM_2022}. For an $E_{2g}$-symmetric lattice distortion the detwinning effect should be effective in our configuration, with the [100] crystallographic axis along the poling direction. However, as shown in Fig. \ref{FigResistanceGlued}b, we do not detect any difference between the transverse and longitudinal resistances. Down to the lowest temperature there is minute difference between the normalized resistance in the transverse and longitudinal channels. This absence of resistance anisotropy in the presence of finite $\epsilon_{\rm{xx}}-\epsilon_{\rm{yy}}$ strain is another evidence for the lack of $E_{2g}$ electronic nematicity, or any rotational symmetry breaking, in direct support of our conclusions.   \\


\subsection{Strain sweeps at fixed temperatures}

\begin{figure*}[h]
\centering
\includegraphics[width=17.2cm]{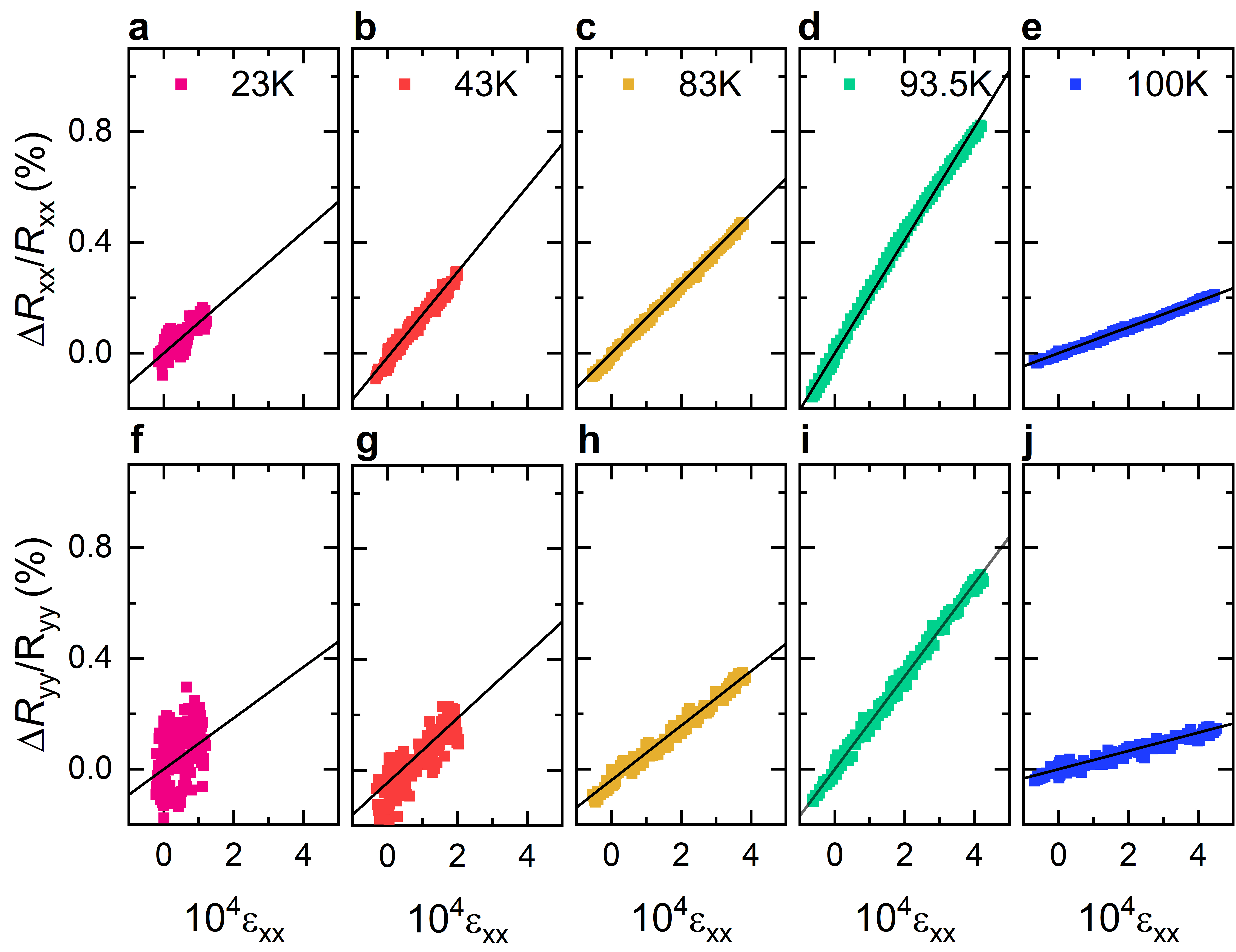}
\caption{\textbf{Resistance variation with strain across $T_{\rm{CDW}}$. (a-e)} Longitudinal resistance variation, $(\Delta R/R)_{\rm{xx}}$, as a function of $\epsilon_{\rm{xx}}$ and for the indicated temperatures. The lines corresponds to linear fits of the data whose slopes equal to $1/R_{\textrm{xx}}dR_{\textrm{xx}} /d\epsilon_{\rm{xx}}$, i.e. the quantity reported in Fig. 2a of the main manuscript. In \textbf{(f-j)} the corresponding transverse measurements, within the same sample, are presented. Note that in all panels we report raw data acquired during a warming cycle.}
\label{FigStrainSweep}
\end{figure*}

We now turn to the measurements acquired at fixed temperatures while sweeping the voltage of the piezoelectric, i.e. the type of measurements reported in the main text. Resistance versus strain sweeps are reproduced in Fig. \ref{FigStrainSweep} at a few selected temperatures, both in the longitudinal and transverse channels as measured on the very same sample, without unmounting it from the piezoelectric stack. First, it is noticeable that at all temperatures the longitudinal and transverse elastoresistances are positive. This clearly contrasts with established electronic nematic systems as, e.g., BaFe$_2$As$_2$ \citep{Chu_Science_2012}. This directly implies that the isotropic $A_{\rm{1g}}$ elastoresistance coefficient is larger than the anisotropic $E_{\rm{2g}}$ one. Second, it is insightful to inspect the magnitude of the effect. At most, i.e. at \Tcdw~where the elastoresistance shows a sharp peak, the resolved resistance variation are less than 1\%, well below the values measured in BaFe$_2$As$_2$ or its nickel analogue  BaNi$_2$As$_2$ using the very same setup \citep{Frachet_npjQM_2022}.\\

\section{Evidence for CDW fluctuations}

\begin{figure*}[h]
\centering
\includegraphics[width=10.2cm]{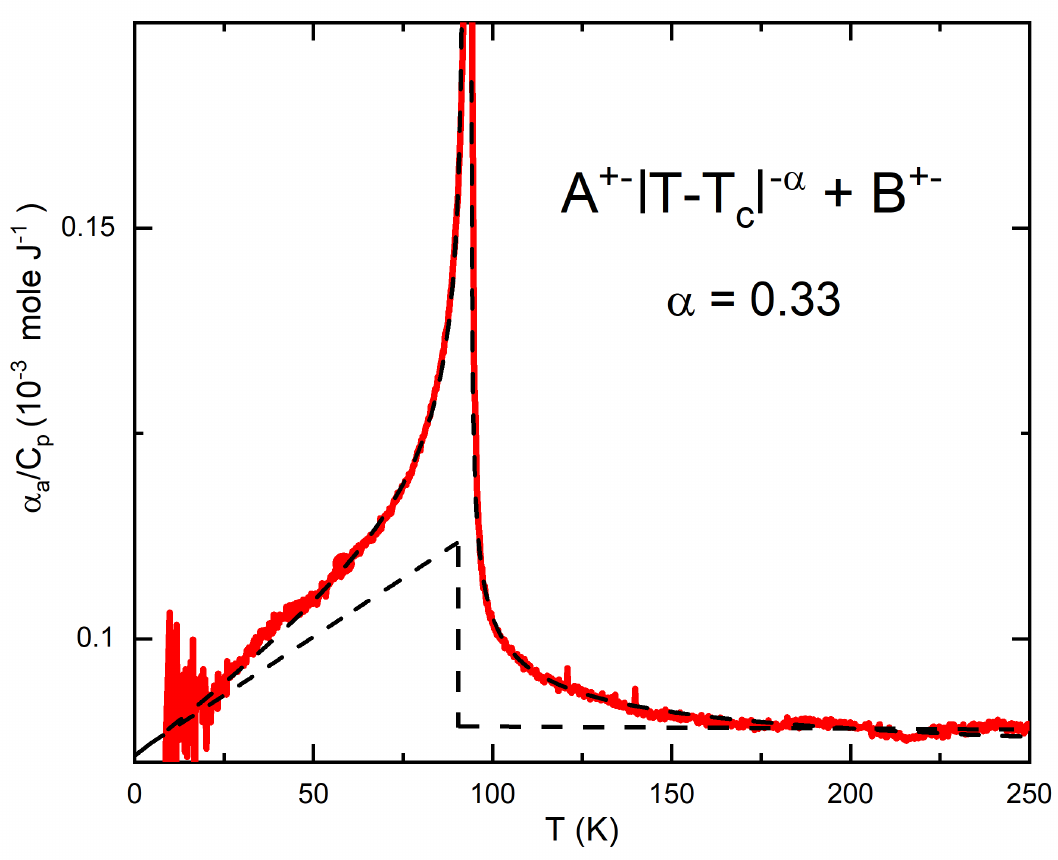}
\caption{\textbf{CDW fluctuations.} In-plane thermal expansion $\alpha_a$(T) divided by the heat capacity $C_p$(T) near the CDW transition, which is representative of a uniaxial Grüneisen parameter related to the pressure dependence of the CDW. The dashed-line discontinuity represents the mean-field jump. The black continuous line shows a fit to the data with a critical exponent $\alpha$ = 0.33.}
\label{Fluctuations}
\end{figure*}

Although the CDW transition is clearly of first-order, Fig. \ref{Fluctuations} demonstrates the presence of significant CDW fluctuations both above and below the transition in an appropriate Grüneisen parameter. Here, we have plotted the ratio of the in-plane thermal expansion $\alpha_a$(T) divided by the heat capacity $C_p$(T), which is representative of a uniaxial Grüneisen parameter related to the pressure dependence of the CDW. Besides a nearly constant background due to the phonon anharmonicity, this quantity is very reminiscent of a critical second-order transition with large fluctuation effects on top of the anharmonic phonon background. In fact, it is possible to describe the data quite well with the usual expression for critical fluctuations with a critical exponent of $\alpha$ = 0.33 over a surprisingly large temperature range (black dashed line in Fig. \ref{Fluctuations}).  Here, we have added a mean-field jump at T$_{CDW}$ below the transition and have also used slightly different T$_{CDW}$s above and below the transition, which is necessary due to the first-order nature of the transition.  Besides providing evidence for these significant fluctuation effects, Fig. \ref{Fluctuations} again demonstrates the absence of any other thermodynamic anomalies above or below T$_{CDW}$ with higher resolution than in the main Figures.

\bibliography{apssamp}